\documentclass
[superscriptaddress,secnumarabic,amssymb,amsmath,nobibnotes,aps,prd,showkeys,showpacs,nofootinbib,onecolumn,notitlepage]{revtex4}%
\usepackage{setspace}
\usepackage{color}
\usepackage{amsmath}
\usepackage{amsfonts}
\usepackage{verbatim}
\usepackage{amssymb}
\usepackage{graphicx,bm}
\usepackage{graphicx}
\usepackage{bbm}
\usepackage{amsmath}
\usepackage{amssymb}
\usepackage{bbm}
\usepackage{amssymb}
\usepackage{graphicx,bm}
\usepackage{graphicx}
\usepackage{bbm}
\usepackage{epstopdf}
\usepackage[shortlabels]{enumitem}
\usepackage[caption=false]{subfig}%
\providecommand{\U}[1]{\protect\rule{.1in}{.1in}}

\newcommand{\be}{\begin{equation}}
\newcommand{\ee}{\end{equation}}

\newcommand{\mincir}{\raise
-3.truept\hbox{\rlap{\hbox{$\sim$}}\raise4.truept\hbox{$<$}\ }}
\newcommand{\magcir}{\raise
-3.truept\hbox{\rlap{\hbox{$\sim$}}\raise4.truept\hbox{$>$}\ }}

\ifx\pdfoutput\relax\let\pdfoutput=\undefined\fi
\newcount\msipdfoutput
\ifx\pdfoutput\undefined\else
\ifcase\pdfoutput\else
\ifx\paperwidth\undefined\else
\ifdim\paperheight=0pt\relax\else\pdfpageheight\paperheight\fi
\ifdim\paperwidth=0pt\relax\else\pdfpagewidth\paperwidth\fi
\fi\fi\fi
\begin{document}
\title{Crossing the phantom divide line as an effect of quantum transitions}
\author{N. Dimakis}
\email{nsdimakis@scu.edu.cn ; nsdimakis@gmail.com}
\affiliation{Center for Theoretical Physics, College of Physics Sichuan University, Chengdu
610064, China}
\author{Andronikos Paliathanasis}
\email{anpaliat@phys.uoa.gr}
\affiliation{Institute of Systems Science, Durban University of Technology, PO Box 1334,
Durban 4000, South Africa}

\begin{abstract}
We consider the Chiral cosmological model consisting of two scalar fields
minimally coupled to gravity. In the context of a
Friedmann--Lema\^{\i}tre--Robertson--Walker (FLRW) spacetime, and for massless
fields in the presence of a cosmological constant, we present the general
solution of the field equations. The minisuperspace configuration that
possesses maximal symmetry leads to scenarios which - depending on the
admissible value of the parameters - correspond to a quintessence, quintom or
phantom case. The canonical quantization of the model retrieves this
distinction as different families of quantum states. The crossing of the
phantom line is related to the existence of free or bound states for the
Casimir operator of the symmetry algebra of the fields. The classical
singularity, which is present in the quintessence solution, is also resolved
at the quantum level.

\end{abstract}
\keywords{Scalar field; cosmology; exact solutions; quantum cosmology}\maketitle
\date{\today}

\section{Introduction}

Scalar fields play an important role in cosmological studies because they
provide processes with the means to explain the recent cosmological
observations \cite{Teg,Kowal,Komatsu,planck,planck18}. Specifically, the
obtained cosmological data indicate that our universe has gone through two
acceleration phases in its evolutionary history. The late accelerated
expansion, which appears to continue until today, and an early acceleration
phase known as inflation.

In order to expain the observed isotropization of the universe at large scales
it has been suggested that the universe has gone through an expansion era
known as inflation. The main theoretical mechanism which has been proposed to
describe the inflationary era is based on the existence of a scalar field
known as the inflaton \cite{guth}. The latter dominates for a short period the
dynamics which drive the evolution of the universe so that, at large scales,
it appears to be isotropic and homogeneous.

For the recent acceleration era of the universe, scalar fields can also assume
the role of the dark energy \cite{Ratra,BarrowSF}. To this end, cosmologists
consider a contribution in the Einstein field equations with the property that
the resulting effective fluid has an equation of state parameter $w<-\frac
{1}{3}$. The pure cosmological constant $\Lambda$ model is the simplest dark
energy model in terms of dynamics and its behaviour. The corresponding
equation of state parameter for the fluid source described by the cosmological
constant is $w_{\Lambda}=-1$ and does not vary during the evolution of the
universe. However, because of its simplicity in terms of dynamics, the
cosmological constant cannot explain the complete cosmological history and
suffers from two major drawbacks, known as the fine tuning and the coincidence
problem \cite{Weinberg1,Padmanabhan1}. Consequently, the models proposed by
cosmologists to overpass these difficulties, made the scalar field important
in the study of the cosmological evolution.

Recently in \cite{bavacu}, the authors used a classical scalar field, called
the \textquotedblleft vacuumon", in the description of running vacuum models.
It was demonstrated that the scalar field description is very helpful for the
explanation of the physical mechanisms of the running vacuum models during
both the early universe and the late time cosmic acceleration \cite{bavacu}.
This was not the first attempt where scalar fields were used for the depiction
of various dark energy models: In \cite{sol1v} a scalar field configuration
was used as a representative model for the description of a running vacuum
theory. The scalar field representation of a matter creation model was
presented in \cite{spn1}, while a realization of the generalized Chaplygin gas
was given in \cite{cg1}; we refer the reader in \cite{inf1} and references
therein for a scalar field description of the modified Chaplygin gas.

Furthermore, scalar fields can be used as new degrees of freedom to obtain
analogous results to those obtained from modified theories of gravity, which
generalize Einstein's theory of General Relativity
\cite{bd,fr1,fr2,fr3,ft1,ft2,fg1,fg2,Oik}. Indeed, the supergravity
inflationary model $R+qR^{2}$ \cite{aastar}, known as Starobinsky inflation
can be described by a scalar field, where the potential of the inflaton is
constructed by the supergravity model by using a Lagrange multiplier and a
conformal transformation \cite{oo0}; a standard approach which relates the
Jordan and the Einstein frames of scalar field models \cite{oo1,oo2,oo3,oo4}.

Quintessence is the simplest scalar field proposed by Ratra et al.
\cite{Ratra}. The energy density of the scalar field is always positive while
the parameter for the equation of state, namely $w_{Q}$, is bounded as
$\left\vert w_{Q}\right\vert \leq1$. In the limit $w_{Q}=1$, only the kinetic
part of the scalar field contributes in the evolution of the universe. On the
other hand, when $w_{Q}=-1$, the scalar field mimics the cosmological
constant. One of the main characteristics of the quintessence is the unstable
tracker solution, for more details on the dynamics of the quintessence we
refer the reader in \cite{qui1,qui2,qui3,qui4,qui5,qui6,qui7}.

The cosmological observations does not provide a lower limit for the value of
$w$. While it has a value close to $-1$, the values in the range $w<-1$ are
not excluded by the cosmological observations \cite{ph1,ph2,ph3,ph4,ph5,ph6}.
Hence, scalar fields with phantom kinetic energy have been introduced in the
literature \cite{sph1,sph2}. The phantom cosmological models generally lead to
a big rip, however - as it has been shown by various studies - this can be
overpassed, for details see \cite{br01,br02,br03,br04}.

It has been proposed that the equation of state parameter may have crossed the
phantom divide line more than once, leading to the quintom scalar field
cosmology \cite{qq1,qq2,qq3}. The latter theory consists of two-scalar fields
(one quintessence and one phantom) which interact, not necessarily in the
potential term. In quintom cosmology, the second scalar field introduces new
degrees of freedom, which provide more possibilities towards the cosmological
evolution. Therefore, multi-scalar field models have been introduced by
cosmologists for the description of the various eras of the universe
\cite{mmf1,mmf2,mmf4}.

In this work, we are interested in a two-scalar field cosmology known as
Chiral cosmology \cite{chir3}. In this gravitational theory, the two scalar
fields are minimally coupled to gravity; however they are necessarily
interacting in the kinetic part. This cosmological model is related to the
non-linear sigma cosmological model \cite{sigm0,sigm1}. This specific theory
is also linked to the $\alpha-$attractor model which has been used as an
alternative for the description of inflation and also as a dark energy model
\cite{al1,al2,al3,al4,al5,al6,all7,al8}.

For the latter gravitational theory and for a homogeneous
Friedmann--Lema\^{\i}tre--Robertson--Walker (FLRW) spacetime we present a set
of analytic solutions which can describe the phantom and the quintessence
epochs of the universe. Additionally, we demonstrate how these two distinct
cosmological solutions correspond to different sets of quantum states. Thus,
making the passing of the phantom divide line a matter of quantum transitions
(e.g. bound to free states).

Specifically, because the gravitational theory of our consideration admits a
minisuperspace description, we apply the canonical quantization which leads to
the Wheeler-DeWitt equation \cite{DeWitt}. At the quantum level we define,
from the classical conservation laws of the field equations, quantum operators
which are used as supplementary conditions over the Wheeler-DeWitt equation.
This approach has been applied before and has lead to various interesting
results in cosmological and gravitational models \cite{qu2,qu3,Pailas} where
it can be seen that in the semiclassical limit the curvature singularities can
be avoided \cite{qu4}. The novelty of this approach is that quantum
observables and their eigenvalues can be related to classical constants of
integration appearing in the metric \cite{qu5,qu6,qu7}. In this work, we see
that the Chiral cosmology, which classically may lead to either quintessence
or a phantom field(s), at the quantum level brings about different sets of
eigenvalues depending on the classical equivalent of the system. Thus
connecting different quantum states to distinct classical behaviours.

In Section \ref{sec2}, we present the cosmological model of our consideration
which is that of Chiral cosmology in a FLRW spacetime. For the scalar field
potential we consider the simplest case, which requires the latter to be
constant, such that only the kinetic part of the scalar fields is a
time-varying function. In other words, we assume the two scalar fields to be
massless. We calculate the conservation laws of the field equation which are
generated by the elements of the $\mathfrak{so}(1,2)$ Lie algebra. The
classical solution of the field equations for the cases with or without
spatial curvature is presented in Section \ref{sec3}. For the spatially flat
spacetime and for a specific value of one of the free parameters of the model
we are able to write the analytic solution in closed form functions. We
observe that for different values of the integration constants we are able to
recover distinct solutions which we call quintessence ($w>-1$) or phantom
($w<-1$) epochs.

The quantization method that is followed is presented in Section \ref{sec4}.
We calculate the wave function of the universe for the generic model of our
consideration. In Section \ref{sec5} we present, for the specific values of
the parameters examined at the classical level, an extended analysis where we
show how the two different classical behaviours, i.e. the quintessence and the
phantom epochs, are represented by different families of quantum states.
Finally, our discussion on the results in given in Section \ref{sec6} and in
the appendices some mathematical calculations which are necessary for our
analysis are presented.

\section{The cosmological model}

\label{sec2}

We consider two massless scalar fields minimally coupled to Einstein's gravity
in the presence of a cosmological constant $\Lambda$
\begin{equation}
S=\int\sqrt{-g}\left[  \frac{1}{2}R-\Lambda-\frac{1}{2}\left(  \nabla_{\kappa
}\phi\nabla^{\kappa}\phi+\sinh^{2}\left(  \lambda\phi\right)  \nabla_{\kappa
}\chi\nabla^{\kappa}\chi\right)  \right]  d^{4}x, \label{action}%
\end{equation}
with the two scalar fields, $\phi$ and $\chi$, interacting in the kinetic
part. The two-dimensional space, which is defined by the kinetic term of the
scalar fields, is a space of constant curvature. Such an Action Integral is
invariant under the $SO(1,2)$ group \cite{Maharana1,Maharana2,Cordero}. The
$\lambda$ appearing in \eqref{action} is a nonzero free parameter associated
to the scalar constant curvature of the two-dimensional field space.

Variation of the Action Integral (\ref{action}) with respect to the metric
tensor and the scalar fields leads to the gravitational field equations which
are:
\begin{align}
R_{\mu\nu}-\frac{1}{2}Rg_{\mu\nu} +\Lambda g_{\mu\nu}  &  =T_{\mu\nu
},\label{fe1}\\
\nabla_{\kappa}\nabla^{\kappa}\phi-\frac{\lambda}{2}\sinh\left(  2\lambda
\phi\right)  \nabla^{\kappa}\chi\nabla_{\kappa}\chi &  =0,\label{fe2}\\
\nabla_{\kappa}\left(  \sinh^{2}\left(  \lambda\phi\right)  \nabla^{\kappa
}\chi\right)   &  =0, \label{fe3}%
\end{align}
where $R$, $R_{\mu\nu}$ are the Ricci scalar and tensor corresponding to the
spacetime metric $g_{\mu\nu}$. Moreover, the energy momentum tensor $T_{\mu
\nu}~$consisted by the two scalar fields is given by the following formula:
\begin{equation}
T_{\mu\nu}=\nabla_{\mu}\phi\nabla_{\nu}\phi-\frac{1}{2}g_{\mu\nu}%
\nabla^{\kappa}\phi\nabla_{\kappa}\phi+\sinh^{2}\left(  \lambda\phi\right)
\left[  \nabla_{\mu}\chi\nabla_{\nu}\chi-\frac{1}{2}g_{\mu\nu}\nabla^{\kappa
}\chi\nabla_{\kappa}\chi\right]  . \label{Tmunu}%
\end{equation}

We assume that the spacetime is described by the FLRW line element
\begin{equation}
ds^{2}=-N(t)^{2}dt^{2}+\frac{a(t)^{2}}{1-kr^{2}}\left(  dr^{2}+r^{2}%
d\theta^{2}+r^{2}\sin^{2}\theta d\varphi^{2}\right)  , \label{FLRW}%
\end{equation}
where $k=-1,0,+1$ denotes the spatial curvature. Unless we want the fields
$\phi$ and $\chi$ to have mutually cancelling contributions in \eqref{action},
leading to $T_{\mu\nu}\equiv0$, we need to set (for the consistency of
Einstein's equations) $\phi=\phi(t)$ and $\chi=\chi(t)$.

By freezing out the spatial coordinates in the original action \eqref{action}
we are led to a minisuperspace Lagrangian which reads
\begin{equation}
L=\frac{1}{2N}\left[  a^{3}\left(  \dot{\phi}^{2}+\sinh^{2}\left(  \lambda
\phi\right)  \dot{\chi}^{2}\right)  -6a\dot{a}^{2}\right]  +N\left(
3ka-\Lambda a^{3}\right)  . \label{Lag}%
\end{equation}
The Euler-Lagrange equations are generated by varying Lagrangian (\ref{Lag})
with respect to the kinematic quantities $\left\{  N,a,\phi,\chi\right\}  $,
and can be expressed as%
\begin{equation}
\frac{6a\dot{a}^{2}}{N}-\frac{a^{3}\dot{\phi}^{2}}{N}-\frac{a^{3}\dot{\chi
}^{2}\sinh^{2}(\lambda\phi)}{N}-2N\Lambda a^{3}+6Nka=0, \label{eulcon}%
\end{equation}%
\begin{align}
\frac{2}{N}\frac{d}{dt}\left(  \frac{\dot{a}}{Na}\right)  +\frac{3\dot{a}^{2}%
}{N^{2}a^{2}}+\frac{\dot{\phi}^{2}}{2N^{2}}+\sinh^{2}(\lambda\phi)\frac
{\dot{\chi}^{2}}{2N^{2}}+\frac{k}{a^{2}}-\Lambda &  =0,\label{euleqchi}\\
\left(  \frac{\dot{N}}{N}-\frac{3\dot{a}}{a}\right)  \dot{\phi}+\frac{\lambda
}{2}\sinh(2\lambda\phi)\dot{\chi}^{2}-\ddot{\phi}  &  =0,\label{euleqchi2}\\
\frac{3\dot{a}\dot{\chi}\sinh^{2}(\lambda\phi)}{a}-\frac{\dot{N}\dot{\chi
}\sinh^{2}(\lambda\phi)}{N}+2\lambda\dot{\chi}\dot{\phi}\sinh(\lambda
\phi)\cosh(\lambda\phi)+\sinh^{2}(\lambda\phi)\ddot{\chi}  &  =0,
\label{euleqchi3}%
\end{align}
these are completely equivalent to the reduced system which is obtained from
the field equations (\ref{fe1})-(\ref{fe3}), under the ansatz \eqref{FLRW} for
the spacetime and a pure time dependence of the fields.

Due to invariance of the action under the $SO(1,2)$ group, the system admits
the point symmetries with generators
\begin{equation}
\xi_{1}=\cos(\lambda\chi)\partial_{\phi}-\coth(\lambda\phi)\sin(\lambda
\chi)\partial_{\chi},\quad\xi_{2}=\sin(\lambda\chi)\partial_{\phi}%
+\coth(\lambda\phi)\cos(\lambda\chi)\partial_{\chi},\quad\xi_{3}%
=\partial_{\chi} \label{sogen}%
\end{equation}
which can be used to construct conservation laws for the field equations by
using the method of variational symmetries. For example, we observe that from
equation (\ref{fe3}) we get the conserved quantity
\begin{equation}
\frac{a^{3}}{N}\sinh^{2}\left(  \lambda\phi\right)  \dot{\chi}=\text{const.},
\end{equation}
which is generated by the symmetry vector field $\xi_{3}$.

In the following section we continue by presenting the analytical solution of
the field equations (\ref{eulcon})-(\ref{euleqchi3}).

\section{Classical solution}

\label{sec3}

For the convenience of our analysis, we perform a reparametrization of the
lapse $N$: $N\mapsto n=2Na^{3}$ that leads to the equivalent point-like
Lagrangian for the field equations
\begin{equation}
\widetilde{L}=\frac{1}{n}G_{\alpha\beta}\dot{q}^{\alpha}\dot{q}^{\beta
}-n\left(  \frac{\Lambda}{2}-\frac{3k}{2a^{2}}\right)  , \label{Lagn}%
\end{equation}
where $q^{\alpha}=(a,\phi,\chi)$. In the case of a spatially flat spacetime
$k=0$, Lagrangian $\widetilde{L}$ describes the motion of a free relativistic
particle of mass $M=\sqrt{\Lambda}$ in the minisuperspace of characterized by
the metric $G_{\mu\nu}$ which reads
\begin{equation}
G_{\alpha\beta}=%
\begin{pmatrix}
-12a^{4} & 0 & 0\\
0 & 2a^{6} & 0\\
0 & 0 & 2a^{6}\sinh^{2}\lambda\phi
\end{pmatrix}
~, \label{minimetric}%
\end{equation}
The corresponding Ricci scalar of the minisuperspace~$G_{\alpha\beta}$ is
calculated to be
\begin{equation}
\mathcal{R}=\frac{3-2\lambda^{2}}{2a^{6}}. \label{miniRicci}%
\end{equation}

The three $\xi_{I}$, $I=1,2,3$, of \eqref{sogen} are Killing vectors of
$G_{\alpha\beta}$. For the particular values $\lambda=\pm\sqrt{\frac{3}{2}}$
the minisuperspace metric $G_{\alpha\beta}$ describes a three dimensional flat
space and admits three additional Killing vectors. This case is the one which
is going to be of most interest in our analysis. Let us begin however with
some general remarks regarding the generic situation.

The three Killing vectors of $G_{\alpha\beta}$ - since they also leave
invariant the potential term of \eqref{Lagn} - define integrals of motion of
the form
\begin{equation}
Q_{i}=\xi_{i}^{\alpha}p_{\alpha}=\xi_{i}^{\alpha}\frac{\partial\widetilde{L}%
}{\partial\dot{q}^{\alpha}},\quad i=1,...,3, \label{clasQ}%
\end{equation}
where $p_{\alpha}=\frac{\partial\tilde{L}}{\partial\dot{q}^{\alpha}}$ are the
momenta associated with the velocities $\dot{a},\dot{\phi}$ and $\dot{\chi}$.

By utilizing the three equations $Q_{i}=\kappa_{i}$, where $\kappa_{i}$ are
constants; it is easy to obtain
\begin{align}
n(t)  &  =\frac{2}{\kappa_{3}}a^{6}\sinh^{2}(\lambda\phi)\dot{\chi},\\
\phi(t)  &  =\frac{1}{\lambda}\coth^{-1}(\alpha\sin(\beta+\lambda\chi)),
\end{align}
where we reparametrized the constants $\kappa_{1}=-2\alpha\kappa_{3}\cos\beta$
and $\kappa_{2}=\alpha\kappa_{3}\sin\beta$ in terms of the new parameters
$\alpha,\beta$.

Substitution of the above expressions into the equations of motion
(\ref{eulcon})-(\ref{euleqchi3}) leaves us only to solve the constraint
\eqref{eulcon}, which reduces to
\begin{equation}
\frac{\dot{\chi}}{\sqrt{6}\kappa_{3}\left(  \alpha^{2}\sin^{2}(\beta
+\lambda\chi)-1\right)  }=\pm\frac{\dot{a}}{a\sqrt{2\Lambda a^{6}%
-6ka^{4}+\left(  \alpha^{2}-1\right)  \kappa_{3}^{2}}}. \label{conrem}%
\end{equation}
The previous equation can be easily integrated to give $a$ in terms of $\chi$.
The latter remains an arbitrary function due to the fact that we did not adopt
some specific time gauge for the system. The resulting expression is quite
complicated and given in terms of an elliptic integral of the third type,
which we refrain from giving here. Nevertheless, in the particular cases where
$\alpha=\pm1$, $\Lambda=0$ and $k=0$ the solution can be written in terms of
elementary functions:

(I) When $\alpha=\pm1$, equation \eqref{conrem} results in
\begin{equation}
\chi(a)=\frac{1}{\lambda}\tan^{-1}\left[  c_{1}\mp\frac{\kappa_{3}\lambda
\sqrt{\Lambda a^{2}-3k}\left(  3k\sqrt{3-\frac{\Lambda a^{2}}{k}}+\sqrt
{3}\Lambda a^{2}\tanh^{-1}\left(  \sqrt{1-\frac{\Lambda a^{2}}{3k}}\right)
\right)  }{6k^{2}a^{2}\sqrt{9-\frac{3\Lambda a^{2}}{k}}}\right]  ,
\end{equation}
where with $c_{1}$ we denote the constant of integration.

(II) The case $\Lambda=0$ leads to
\begin{equation}
a(\chi)=\pm\left[  \frac{\kappa_{3}^{2}}{6k}\left(  \alpha^{2}-1\right)
\text{sech}^{2}\left(  \sqrt{\frac{2}{3}}\frac{1}{\lambda}\tanh^{-1}\left(
\sqrt{\alpha^{2}-1}\tan(\lambda\chi+\beta)\right)  \right)  +c_{1}^{\prime
}\right]  ^{1/4}.
\end{equation}
where $c_{1}^{^{\prime}}$ is the constant of integration.

(III) Finally, when $k=0$ the resulting expression for the scale factor is
\begin{equation}
a(\chi)=c_{2}^{1/3}\left[  \sinh\left(  \sqrt{\frac{3}{2}}\frac{1}{\lambda
}\tanh^{-1}\left(  \sqrt{\alpha^{2}-1}\tan(\lambda\chi+\beta)\right)  +\bar
{c}_{1}\right)  \right]  ^{-1/3}, \label{akzero}%
\end{equation}
where the constant $c_{2}=\sqrt{\frac{(\alpha^{2}-1)\kappa_{3}^{2}}{2\Lambda}%
}$ is introduced and $\bar{c}_{1}$ is again the integration constant.

\subsection{The spatially flat universe with $\lambda=\pm\sqrt{\frac{3}{2}}$}

We mentioned that, for the spatially flat universe, i.e. $k=0$, and when
$\lambda=\pm\sqrt{\frac{3}{2}}$, the resulting Lagrangian \eqref{Lagn}
describes a free relativistic particle moving in a three dimensional flat
space. Thus, for these values of the parameters the system admits three
additional linear in the momenta integrals of motion\footnote{The
generators of the three additional conserved charges are those of the typical
translations, when the minisuperspace is in coordinates where $G_{\alpha\beta
}= \mathrm{diag}(-1,1,1)$.} on top of the conserved charges \eqref{clasQ}
which are produced by the $SO(1,2)$ group whose generators are given by
\eqref{sogen}. In this sense it is a system of maximal symmetry in what
regards the minisuperspace description.

Depending on the admissible values of the parameters involved in the solution,
various gravitational behaviours can be obtained. It is possible to
reparametrize the constants of integration so that the solution reads (for the detailed derivation please see the appendix \ref{app0})%

\begin{equation}
a(t)=a_{0}\left(  \frac{\sin(t+\beta)}{\cos t}\sqrt{1-\frac{1}{\alpha^{2}%
\sin^{2}(t+\beta)}}\right)  ^{1/3}, \label{asol1}%
\end{equation}%
\begin{equation}
N(t)=\pm\left(  \sqrt{3}\alpha\gamma\sin(t+\beta)\cos t\sqrt{1-\frac{1}%
{\alpha^{2}\sin^{2}(t+\beta)}}\right)  ^{-1}, \label{Nsol}%
\end{equation}%
\begin{equation}
\phi(t)=\pm\sqrt{\frac{2}{3}}\coth^{-1}(\alpha\sin(t+\beta)),\quad\chi
(t)=\pm\sqrt{\frac{2}{3}}t, \label{fsol}%
\end{equation}
where we have chosen the time gauge in which essentially the $\chi$ field
becomes the time variable.

Note that the above form of solution is minimal in the sense that we have
eradicated integration constants that can be absorbed by means of
reparametrizations and diffeomorphisms (for details see appendix \ref{app0}).
The constant $a_{0}$ in \eqref{asol1} is of course absorbable by a constant
scaling in the $r$ variable in the metric \eqref{FLRW}. However we choose to
keep it because it can serve to maintain $a(t)$ real under different choices
for the rest of the parameters.

Expressions (\ref{asol1})-(\ref{fsol}) satisfy the field equations
\eqref{euleqchi}-\eqref{euleqchi3}, while the constraint equation
\eqref{eulcon} yields the following relation among constants
\begin{equation}
\gamma^{2}\left(  \alpha^{2}\cos^{2}(\beta)-1\right)  =\Lambda
\label{constcon2}%
\end{equation}

As we may observe from (\ref{asol1})-(\ref{fsol}), the solution is periodic
and the region of $t$ for which you can have real $a(t)$ and $N(t)$ depends on
the free parameters of the solution. We note that the parameters involved in
\eqref{asol1} and \eqref{Nsol} may assume any value as long as the end result
has some real domain of definition. For example, it can be easily seen, that
we can set $\alpha$ and $\gamma$ to be simultaneously imaginary and have both
\eqref{asol1} and \eqref{Nsol} as real functions for some interval of $t$. In
the following section we treat separately two distinct cases which give
interesting behaviours that are related with the rate of the expansion of the universe.

At this point, it is necessary to mention that for the Action Integral of the
form (\ref{action}) with various forms of the potential $V\left(  \phi
,\chi\right)  $, exact and analytical solutions have been found previously in
the literature in \cite{anal1,anal2}.

\subsection{The quintessence\ epoch}

Here we study the solution (\ref{asol1})-(\ref{fsol}) when the related
parameters assume such values, so that it describes a universe whose equation
of state parameter, $w=\frac{P_{\text{eff}}}{\rho_{\text{eff}}}$, of the
relevant effective fluid ranges from $1$ to $-1$. Thus, characterizing what we
shall refer to as a \textquotedblleft quintessence epoch". The
$\rho_{\text{eff}}$ and $P_{\text{eff}}$ are the energy density and the
pressure of the effective cosmological perfect fluid that produces the same
energy-momentum tensor as the $T_{\mu\nu}~$ of (\ref{Tmunu}).

Let us study the functional behaviour of the solution assuming $a_{0}=1$,
$0\leq\beta<\frac{\pi}{2}$ and $\Lambda>0$, while $\alpha,\gamma$ are both
real.\footnote{The range of values of $\beta$ and $\Lambda$ have been assumed
in this manner so that we obtain an expansive behaviour for the scale factor.}
In order for the latter to be true - and given the restriction we set on
$\beta$ - we need to have $|\alpha|>(\cos\beta)^{-1}$. Under these conditions
and assuming $t>0$, \eqref{asol1} and \eqref{Nsol} are real in the interval
$t\in\left(  \sin^{-1}\left(  \frac{1}{|\alpha|}\right)  -\beta,\frac{\pi}%
{2}\right)  $. A behaviour that is being repeated with a period of $\pi$.

\begin{figure}[ptb]
\textbf{ \includegraphics[width=0.8\textwidth]{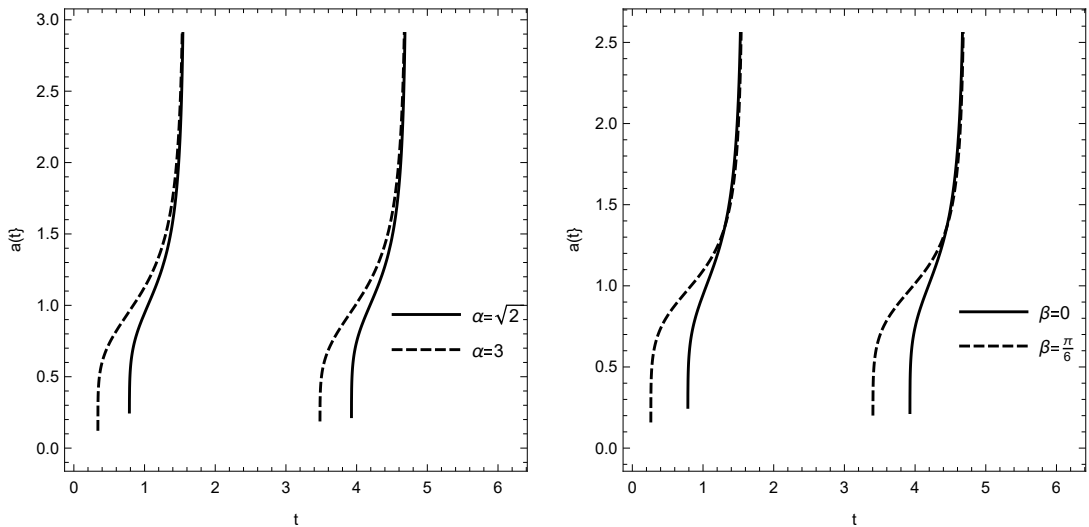} \newline%
}\caption{Qualitative evolution of the scale factor \eqref{asol1} as a
function of the periodic variable $t$. Left Fig. for $\beta=0$ and different
values of $\alpha$, while right Fig. is for $\alpha=\sqrt{2}$ and two
different values of parameter $\beta$. }%
\label{Fig101}%
\end{figure}

\begin{figure}[ptb]
\textbf{ \includegraphics[width=1\textwidth]{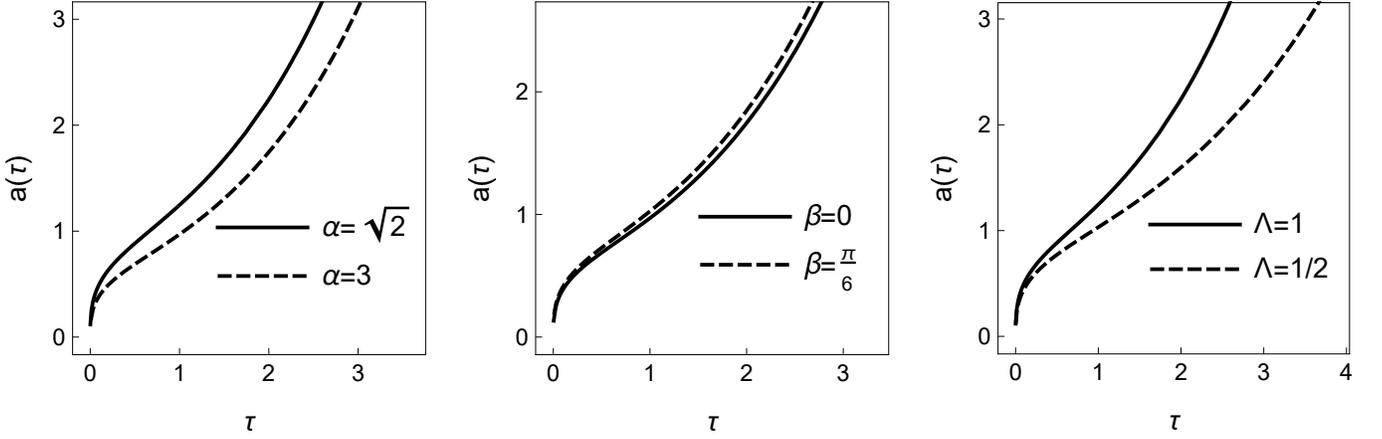} \newline%
}\caption{Qualitative evolution of the scale factor \eqref{asol1} as a
function the proper time $\tau$. Left Fig. is for $\beta=0,~\Lambda=1$ and
different values of $\alpha$. Middle Fig. is for $\alpha=3,~\Lambda=1$ and
different values of $\beta$, while right Fig. if different values of $\Lambda$
where $\alpha=\sqrt{2}$ and $\beta=0$. }%
\label{Fig20}%
\end{figure}

\begin{figure}[ptb]
\centering\textbf{ \includegraphics[width=0.4\textwidth]{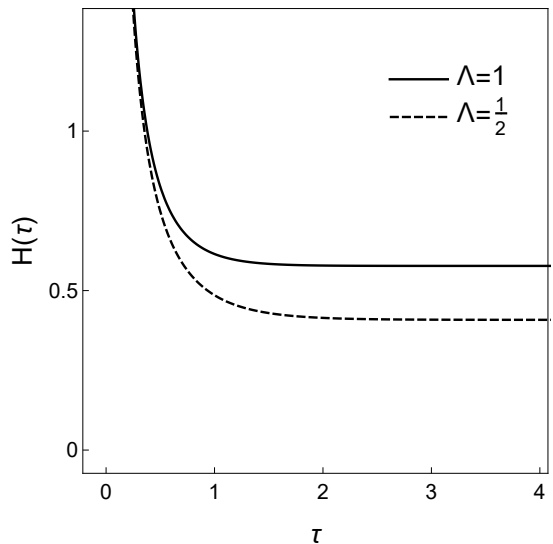}
}\caption{Qualitative behaviour of the Hubble function $H(\tau)$ for two
different values of $\Lambda$ $\ $and $\alpha=\sqrt{2},\beta=0$.}%
\label{Fig30}%
\end{figure}

In figure \ref{Fig101} we demonstrate how the scale factor $a\left(  t\right)
$ is affected by different values of the parameters $\alpha$ and $\beta$. In
order to obtain a more physical insight of the solution we need to express the
result in terms of the cosmological (or cosmic) time $\tau$ in which
$N(\tau)=1$. Thus, we introduce a new time variable $\tau$ given by
\begin{equation}
\tau(t)=\int\!\!N(t)dt . \label{tautot}%
\end{equation}
The application of its inverse however in the analytic solution of the
problem, cannot in general result in expressions given in terms of elementary
functions. Nevertheless, we are able, in figures \ref{Fig20} and \ref{Fig30}
to give some parametric plots for the scale factor $a(\tau)$ and the Hubble
function $H(\tau)=\frac{1}{a\left(  \tau\right)  }\frac{da\left(  \tau\right)
}{d\tau}$ as functions of the cosmological time $\tau$ defined by
\eqref{tautot}. We need to note that in the plots, the choice of the values of
the parameters is made so as to demonstrate in a simple manner how they affect
the behaviour of the functions, it is not with reference to observational values.

As we see in figure \ref{Fig20} the values of $\alpha$ and $\Lambda$ affect
the \textquotedblleft steepness" of the expansion in an opposite manner.
Larger values of $\alpha$ lead to a milder expansion, while in what regards
$\Lambda$ this happens for smaller values. On the other hand $\beta$ just
translates the graph in time, which is unimportant if you consider that
transformation \eqref{tautot} already has the freedom of adding an arbitrary
constant and thus shifting in time the whole graph. In figure \ref{Fig30} we
see how the Hubble parameter is affected by $\Lambda$, which dominates the
value towards which $H(\tau)$ asymptotically tends. Recall that in the time
gauge of \eqref{asol1} and \eqref{Nsol} the Hubble function is given by
$H(t)=\frac{1}{aN}\frac{da}{dt}$.

It is interesting to note that the finite region $t\in\left(  \sin^{-1}\left(
\frac{1}{|\alpha|}\right)  -\beta,\frac{\pi}{2}\right)  $ corresponds through
the inverse of \eqref{tautot} to a cosmological time $\tau\in(0,+\infty)$.
What it is more, the limit $t\rightarrow\sin^{-1}\left(  \frac{1}{|\alpha
|}\right)  -\beta$ (or equivalently $\tau\rightarrow0$ in cosmological time)
corresponds to a curvature singularity for the spacetime whose metric is
characterized by \eqref{asol1} and \eqref{Nsol}.

The effective fluid energy density $\rho_{\text{eff}}$ and pressure
$P_{\text{eff}}$ that we can read from the energy momentum tensor
\eqref{Tmunu} are
\begin{equation}
\rho_{\text{eff}}=\frac{1}{2N^{2}}\left(  \dot{\phi}^{2}+\sinh^{2}\left(
\lambda\phi\right)  \dot{\chi}^{2}\right)  ,\quad P_{\text{eff}}%
=\rho_{\text{eff}}-2\Lambda.
\end{equation}
In the particular case that we study, where $\lambda=\pm\sqrt{\frac{3}{2}}$
and the analytical solution is given by \eqref{asol1} under the constraint
\eqref{constcon2}, we calculate the equation of state function $w$ to be
\begin{equation}
w(t)=\frac{P_{\text{eff}}}{\rho_{\text{eff}}}=-1+\frac{8\left(  \alpha
^{2}-1\right)  \cos^{2}t}{\left(  \alpha^{2}\sin(2\beta+t)+\left(  \alpha
^{2}-2\right)  \sin t\right)  ^{2}}. \label{wratio}%
\end{equation}
From (\ref{wratio}) we observe that for $w\left(  \frac{\pi}{2}\right)  =-1$,
that is, only the cosmological constant contributes in the universe in this
limit. However, the latter is not true near the cosmological singularity, i.e.
$t\rightarrow t_{0}=\sin^{-1}\left(  \frac{1}{|\alpha|}\right)  -\beta$. In
this case $w\left(  t_{0}\right)  =1$. We remark that for $\alpha=1,$ the
general solution reduces exactly to that of the de Sitter universe.

\begin{figure}[ptb]
\centering\textbf{ \includegraphics[width=0.5\textwidth]{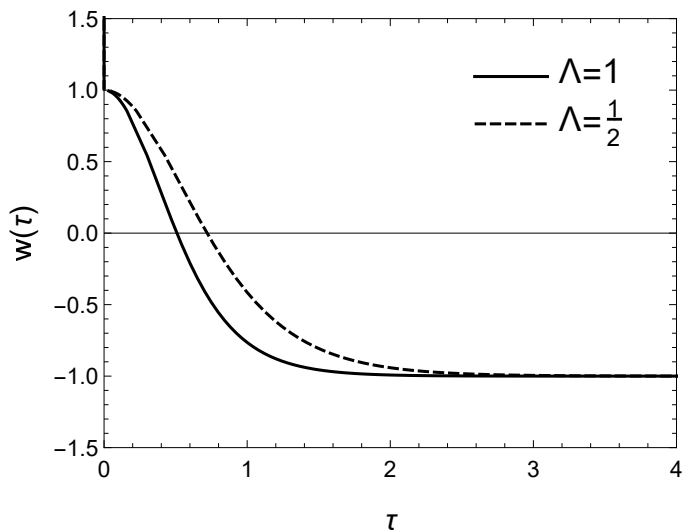}}%
\caption{The parameter for equation of state function $w(\tau)$ in terms of
the cosmological time, for two different values of $\Lambda$ ($\alpha
=3,\beta=0$) for the quintessence\ epoch. We observe that in the limits
function $w\left(  \tau\right)  $ takes the values of the bounds which are $1$
and $-1$. }%
\label{Fig40}%
\end{figure}

Although $w(t)$ does not depend on $\Lambda$, when the function is expressed
in terms of the cosmological time $\tau$, the corresponding $w(\tau)$ does.
Because the transformation \eqref{tautot}, through \eqref{Nsol} and
\eqref{constcon2} involves $\Lambda$ in the process. In this manner we are
able to plot what we see in figure \ref{Fig40}.

\subsection{The phantom epoch ($w<-1$)}

As we previously observed, the gravitational part of the solution remains real
if we take $\alpha$ and $\gamma$ to be imaginary. We thus start by assuming,
$\alpha=\mathrm{i}\,\tilde{\alpha}$ and $\gamma=\mathrm{i}\,\tilde{\gamma}$,
where $\tilde{\alpha},\tilde{\gamma}\in\mathbb{R}$. With the help of this
substitution, the analytic solution (\ref{asol1})-(\ref{fsol}) can be brought
to the form%
\begin{align}
a(t)  &  = a_{0} \left(  \frac{\sin(t+\beta)}{\cos t}\sqrt{1+\frac{1}%
{\tilde{\alpha}^{2}\sin^{2}(t+\beta)}}\right)  ^{1/3},\label{asol2}\\
N(t)  &  =\pm\left(  \sqrt{3}\tilde{\alpha}\tilde{\gamma}\sin(t+\beta)\cos
t\sqrt{1+\frac{1}{\alpha^{2}\sin^{2}(t+\beta)}}\right)  ^{-1},\label{Nsol2}\\
\phi(t)  &  =\pm\mathrm{i}\sqrt{\frac{2}{3}}\cot^{-1}(\tilde{\alpha}%
\sin(t+\beta)),\quad\chi(t)=\sqrt{\frac{2}{3}}t \label{fsol22}%
\end{align}
with the constraint (\ref{constcon2}) among the constants becoming
\begin{equation}
\tilde{\gamma}^{2}=\frac{\Lambda}{\tilde{\alpha}^{2}\cos^{2}(\beta)+1}.
\label{constcon3}%
\end{equation}

We observe that this parameterization results in an imaginary field $\phi$,
which - by looking at the action \eqref{action} - signifies that both $\phi$
and $\chi$ become phantom fields. Of course, we expect that this allows $w(t)$
to cross the phantom divide line and to take values smaller than $-1$. Because
of the latter property, we call that era \textquotedblleft phantom
epoch\textquotedblright.

\begin{figure}[ptb]
\centering\textbf{ \includegraphics[width=1\textwidth]{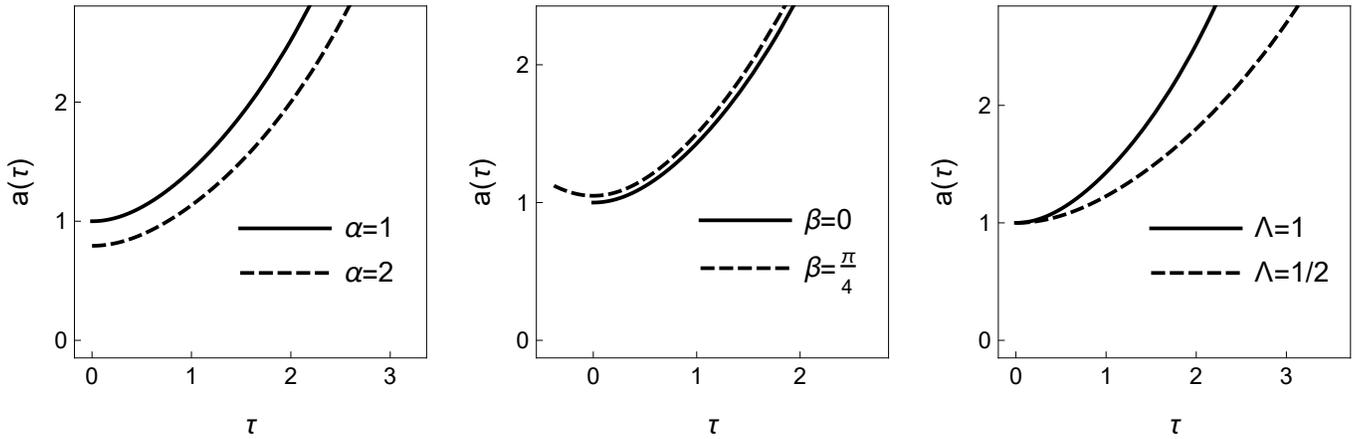} \newline%
}\caption{Qualitative evolution of the scale factor $a\left(  \tau\right)  $
for the phantom epoch. In all plots we assume $a_{0}=1$. Left Fig. is for
$\beta=0,~\Lambda=1$ and for different values of parameter $\alpha$. Middle
Fig. is for $\alpha=1,~\Lambda=1$ and varrying parameter $\beta$. Finally,
right Fig. is for varying parameter $\Lambda$ and $\alpha=1,~\beta=0$. From
the plot we observe that there is not any initial cosmological singularity and
that the universe bounces. }%
\label{Fig50}%
\end{figure}

Let us assume once more a phase $0\leq\beta<\frac{\pi}{2}$. Then, $a(t)$ and
$N(t)$ as given by \eqref{asol2}, \eqref{Nsol2} are real in the region
$t\in[-\beta,\frac{\pi}{2})$, assuming $a_{0}\in\mathbb{R}$, with the same
behaviour being repeated with a period of $\pi$. In figure \ref{Fig50} we
present the plot of the scale factor $a(\tau)$ as function of the cosmological
time $\tau$ defined by \eqref{tautot}. We observe that at the origin, $\tau
=0$, the scale factor obtains a finite non-zero value. It is also easy to
verify, by looking at the Ricci and Kretschmann scalars, $R$ and
$R_{\kappa\lambda\mu\nu}R^{\kappa\lambda\mu\nu}$ respectively, that the
spacetime characterized by \eqref{asol2} and \eqref{Nsol2} has no curvature
singularity. For example, by using \eqref{asol2}, \eqref{Nsol2} and
\eqref{constcon3}, the Ricci scalar becomes
\begin{equation}
R = \frac{\Lambda\left[ 3 \tilde{\alpha} ^{2}+2 \tilde{\alpha}^{2} \cos(\beta)
\sin(\beta+t) \left( \tilde{\alpha}^{2} \sin(2 \beta+t)+\left( \tilde{\alpha}
^{2}+4\right)  \sin(t)\right) +\left( 3\tilde{\alpha}^{2}+1\right)  \cos(2
t)+5\right] }{\left( \tilde{\alpha}^{2} \cos^{2}(\beta)+1\right)  \left(
\tilde{\alpha}^{2} \sin^{2}(\beta+t)+1\right) },
\end{equation}
which is finite for the values of the parameters considered, i.e.
$\tilde{\alpha} \in\mathbb{R}$. The numerator consists of trigonometric
functions which are bounded, while the denominator cannot be zero. The same is
true for the Kretschmann scalar, whose expression however is quite more
complicated and which we avoid to present here. We see thus that in this case
no curvature singularity occurs and, as we demonstrate below, a bounce
solution is obtained.

\begin{figure}[ptb]
\centering\textbf{ \includegraphics[width=0.8\textwidth]{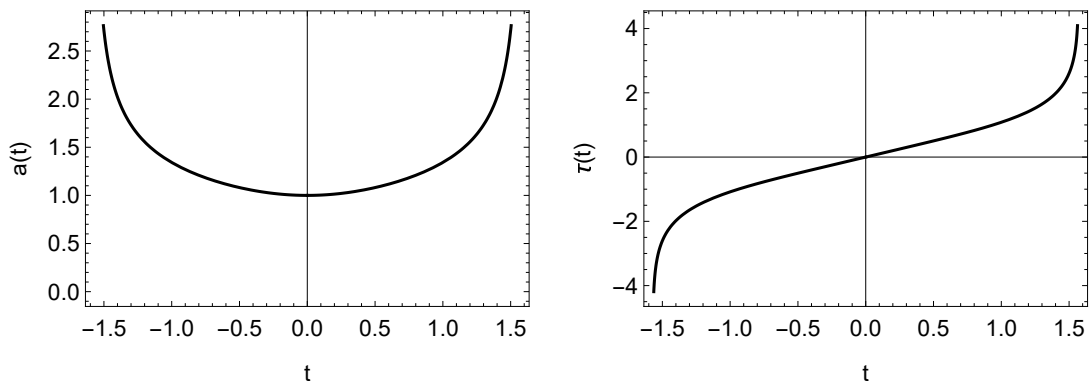}
\newline}\caption{Plots of the scale factor $a(t)$ of \eqref{asolbounce} and
of the cosmological time $\tau(t)$ given by \eqref{cosmict} for the values
$\tilde{\alpha}=1$, $\beta=0$ and $\gamma=\frac{1}{\sqrt{3}}$. The region
$t\in\left( -\frac{\pi}{2},\frac{\pi}{2}\right) $ corresponds to $\tau(t)
\in(-\infty, +\infty)$. The universe bounces at $\tau=0$ and the scale factor
acquires a nonzero minimum value.}%
\label{Fig60}%
\end{figure}

In order to expose the bouncing solution we need to write an extension
of what we see in Fig. \ref{Fig50} for times before $\tau=0$. Due to the
arbitrariness of a multiplicative constant in $a(t)$ and because $N(t)$ is
invariant in an overall sign (see eqs. \eqref{asol2} and \eqref{Nsol2}
respectively) we can construct a smooth extension of the solution for $\tau
<0$. In order to do so we require to have a real expression for the scale
factor throughout the region $t \in\left( -\frac{\pi}{2},\frac{\pi}{2}\right)
$. The latter is achieved by the function 
\begin{equation}
\label{asolbounce}a(t) = \Bigg| \frac{\sin(t+\beta)}{\cos t}\sqrt{1+\frac
{1}{\tilde{\alpha}^{2}\sin^{2}(t+\beta)}} \Bigg|^{1/3} .
\end{equation}
In other words we take $a(t)$ to be constituted of two branches: For
$t\in\left[ -\beta,\frac{\pi}{2}\right)  $ we consider \eqref{asol2} with
$a_{0}=1$, while for $t \in\left( -\frac{\pi}{2}, -\beta\right)  $ we use the
same relation but with $a_{0}=e^{-\mathrm{i} \pi/3}$ (remember that we
consider $\beta\in(0,\frac{\pi}{2})$). Note that expression \eqref{asolbounce}
for the scale factor is continuous at the limit $t \rightarrow- \beta$ and the
same is true for its derivatives with respect to $t$. The latter does not hold
for the expresion without the absolute value. In a similar fashion we need to
construct a smooth expression for $N(t)$. The arbitrariness of the overall
sign in \eqref{Nsol2} is enough for that matter, hence we take
\begin{equation}
\label{Nsolbounce}N(t) =
\begin{cases}
\left(  \sqrt{3}\tilde{\alpha}\tilde{\gamma}\sin(t+\beta)\cos t\sqrt
{1+\frac{1}{\alpha^{2}\sin^{2}(t+\beta)}}\right) ^{-1}, & \mbox{if }
t\in\left[ -\beta,\frac{\pi}{2}\right) \\
-\left(  \sqrt{3}\tilde{\alpha}\tilde{\gamma}\sin(t+\beta)\cos t\sqrt
{1+\frac{1}{\alpha^{2}\sin^{2}(t+\beta)}}\right) ^{-1}, & \mbox{if } t
\in\left( -\frac{\pi}{2}, -\beta\right)  .
\end{cases}
\end{equation}
The cosmological time $\tau$ which is related to $t$ through
\eqref{tautot} is calculated to be
\begin{equation}
\label{cosmict}\tau(t) = \frac{2 \tilde{\alpha} \tanh^{-1}\left[ \frac
{\tilde{\alpha}^{2} \sin(2 \beta+t)+\left( \tilde{\alpha}^{2}+2\right)
\sin(t)}{\left( \tilde{\alpha}^{2} (\cos(2 \beta)+1)+2\right) ^{1/2} \left(
\tilde{\alpha}^{2} (1-\cos(2 (t + \beta)))+2\right) ^{1/2}} \right] }{\sqrt{3}
\tilde{\gamma} \left( \tilde{\alpha}^{2} (\cos(2 \beta)+1)+2\right) ^{1/2}
\left( \tilde{\alpha}^{2} (1-\cos(2 (t + \beta)))+2\right) ^{1/2}} \left(
\frac{1}{\tilde{\alpha}^{2} \sin^{2}(t + \beta)}+1\right) ^{1/2}
\Big|\sin(t+\beta) \Big| .
\end{equation}
For this function, the zero of the cosmological time, $\tau(t_{*})=0$,
is placed at 
\begin{equation}
t=t_{*}:=\tan^{-1}\left( \frac{\tilde{\alpha}^{2} \sin(2 \beta)}{\tilde
{\alpha}^{2} (\cos(2 \beta)+1)+2}\right)
\end{equation}
which is also the time for which the $a(t)$ of \eqref{asolbounce}
assumes its minimum value. It is an easy task to verify that $\dot{a}%
(t_{*})=0$ and $\ddot{a}(t_{*})>0$. As an illustrative example, in Fig.
\ref{Fig60}, we plot the functions \eqref{asolbounce} and \eqref{cosmict} for
some specific values of the parameters.

As far as the scalar fields are concerned regarding the continuity: It is straightforward that $\chi(t)$ of \eqref{fsol22} is continuous since it is effectively the time parameter itself. For $\phi(t)$ in \eqref{fsol22} a similar process as for the $N(t)$ can be followed due to the arbitrariness of the solution in the sign. We can thus take
\begin{equation}
\label{phisolbounce}\phi(t) =
\begin{cases}
\mathrm{i}\sqrt{\frac{2}{3}}\cot^{-1}(\tilde{\alpha}%
\sin(t+\beta)), & \mbox{if }
t\in\left[ -\beta,\frac{\pi}{2}\right) \\
-\mathrm{i}\sqrt{\frac{2}{3}}\cot^{-1}(\tilde{\alpha}%
\sin(t+\beta)), & \mbox{if } t
\in\left( -\frac{\pi}{2}, -\beta\right)  ,
\end{cases}
\end{equation}
which makes the $\phi(t)$ a continuous function throughout $t$. We notice that the $\dot{\phi}$ calculated from \eqref{phisolbounce} possesses a discontinuity in $t\rightarrow -\beta$. Specifically we obtain, $\lim_{t\rightarrow \beta^+} (\dot{\phi} )=-\mathrm{i} \sqrt{\frac{2}{3}} \tilde{a}$ while on the other hand we have $\lim_{t\rightarrow \beta^-} (\dot{\phi} )=\mathrm{i} \sqrt{\frac{2}{3}} \tilde{a}$. However, this creates no particular problem since what enters the Lagrangian and the physically relevant quantities, like the effective energy density $\rho_{eff}$ and the pressure $P_{eff}$ is the $\dot{\phi}^2$ which is continuous in all its domain of definition. The aforementioned discontinuity in $\dot{\phi}$ defined by \eqref{phisolbounce} is present in all odd derivatives of $\phi$, the even derivatives like $\ddot{\phi}$ remain continuous.

From the moment that the scale factor does not become zero and stays
positive we know that non-spacelike geodesics are past-complete \cite{Lam}.
Thus, unlike to what we obtain in the quintessence case, where such a
construction is not possible, here the spacetime has no initial singularity
and describes a bouncing universe. We also note that there is no big rip in
the future since the scale factor does not go to infinity in some finite time
$\tau$. It is for $\tau\rightarrow\pm\infty$ that $a\rightarrow+ \infty$.

In what regards the description of the effective perfect fluid and the
equation of state parameter $w(t)$, we can straightforwardly set $\alpha=
\mathrm{i} \tilde{\alpha}$ in \eqref{wratio} to obtain
\begin{equation}
\label{wratio2}w(t) = -1 -\frac{8 \left(  \tilde{\alpha}^{2}+1\right)
\cos^{2} t}{\left(  \tilde{\alpha}^{2} \sin(2 \beta+t)+\left(  \tilde{\alpha
}^{2}+2\right)  \sin t\right)  ^{2}}%
\end{equation}
which obviously is lesser that $-1$ for all values of $t$. At the limit
$t\rightarrow\frac{\pi}{2}$, the parameter assumes its maximum value
$w_{max}=-1$.

An important remark is that a similar behaviour of expansion for $a(\tau)$ can
be obtained by assuming only $\gamma$ to be imaginary, while $|\alpha|<1$. In
order to keep real the solution expressed by \eqref{asol2} we need to set
$a_{0}=e^{\pm\mathrm{i}\frac{\pi}{6}}$. The plus or minus in the exponent
depends on the branch we need to consider, $t<-\beta$ or $\tau\geq-\beta$
respectively. We can thus state, that $w<-1$ needs $\gamma$ to be imaginary;
from there we have two paths that we may follow depending on the possible
values we may assign to $\alpha$ and $a_{0}$.

It is quite interesting to observe that in the case where we take only
$\gamma$ to be imaginary, the $w<-1$ result is obtained with just one of the
two fields turning phantom, namely the $\chi(t)$. The $\phi(t)$ becomes
complex, but with a constant imaginary part of $\pm\mathrm{i}\frac{\pi}{2}$.
As a result its derivative is real and the corresponding kinetic term in
\eqref{action} does not change sign. On the other hand, the $\sinh^{2}%
(\lambda\phi)$ in front of $\dot{\chi}^{2}$ becomes negative since
$\sinh(x+\mathrm{i}\frac{\pi}{2})=\mathrm{i}\cosh(x)$, which explains how the
line $w(t)=-1$ is crossed.

\subsection{The de Sitter limit}

Until now we studied the behaviour of the classical solution for the field
equations of the model with Action Integral (\ref{action}). We found that
there can be an exact solution which describes a quintessence era and another
solution which describes a bounced universe with at least one phantom field.
While the two models are distinct and they have different analytic solutions
and behaviour, it is important to mention that the two different solutions
have as an attractor the de Sitter universe with $w\left(  t\right)  =-1$.
However, they reach the attractor from different directions. In the
case of the quintessence, $w\left(  t\right) $ decreases and reach
the value $w\left(  t\right)  =-1$, while in the
case of the phantom field, $w\left(  t\right)  $ increases and the
value $w\left(  t\right)  =-1$ plays the role of an upper wall. This
can be seen from equations \eqref{wratio} and \eqref{wratio2}. The limit
$t\rightarrow\frac{\pi}{2}^{-}$ is the asymptotic future of the spacetimes
since in the cosmic time gauge, i.e. when $N=1$, it corresponds to
$\tau\rightarrow+\infty$. In that limit both \eqref{wratio} and
\eqref{wratio2} tend to the value $w=-1$. In the first case the limit is
reached from values $w>-1$, while in the latter from $w<-1$.

Another way to observe the same effect is through the energy momentum
tensor. A simple substitution of either \eqref{asol1}-\eqref{fsol} or
\eqref{asol2}-\eqref{fsol22} leads to all the components of the mixed tensor
$T^{\mu}_{\;\;\nu}$ to be proportional to $\cos^{2}(t)$. As a result, the
$t\rightarrow\frac{\pi}{2}^{-}$ limit leads to $T^{\mu}_{\;\;\nu}=0$ and the
cosmological constant in \eqref{fe1} to dominate the future of both
spacetimes. The de Sitter solution becomes the asymptotic future in both
cases.

Similarly, at the level of the metric, we may consider an expansion of
either solution \eqref{asol1}, \eqref{Nsol} or \eqref{asol2}, \eqref{Nsol2}
when $t\rightarrow\frac{\pi}{2}^{-}$. In both of these cases the leading terms
in the expansion around $\frac{\pi}{2}$ are
\begin{equation}
\label{approxds1}a(t) \propto\left( \frac{1}{\frac{\pi}{2} -t}\right) ^{1/3},
\quad N(t) = \pm\frac{1}{\sqrt{3} \sqrt{\Lambda} \left( \frac{\pi}{2}-t\right)
} ,
\end{equation}
where the relations \eqref{constcon2} and \eqref{constcon3} have also
been applied in each of the corresponding solution. With the help of the
transformation \eqref{tautot} we see that $\tau$ is related to $t$ through
\begin{equation}
t = \frac{1}{2} \left( \pi-e^{\mp\sqrt{3} \sqrt{\Lambda} \tau}\right) .
\end{equation}
We write the resulting approximate line element at the limit
$t\rightarrow\frac{\pi}{2}^{-}$ (or equivalently at $\tau\rightarrow+\infty$)
as
\begin{equation}
ds^{2}_{\tau\rightarrow+ \infty} = -d\tau^{2} + e^{\pm\sqrt{\frac{\Lambda}{3}%
}\tau} \left(  dr^{2} + r^{2} d\theta^{2} + r^{2} \sin^{2}\theta d\varphi^{2}
\right) ,
\end{equation}
which is the de Sitter solution with cosmological constant $\Lambda$.
Note that in the above line element we ignored the multiplicative constant
appearing in $a(t)$ coming from \eqref{approxds1} since it is absorbable with
a diffeomorphism (in particular a constant scaling) in the radial distance
$r$.

Having studied the classical aspects of each case we may now proceed to the
quantum description of the system. In the following section we shall see how
we can pass from one solution to another through quantum processes. The
analysis is performed under the scope of canonical quantum cosmology and the
use of the Wheeler-DeWitt equation.

\section{Quantization of the general model}

\label{sec4}

The quantization of the system described by Lagrangian \eqref{Lagn} is
straightforward. We apply the Dirac-Bergmann \cite{Dirac,AndBer} algorithm for
constrained systems in order to pass to the Hamiltonian formulation.
We sketch the basic steps here, but for more details we refer the
interested reader to textbooks on constrained systems \cite{Dirac2,Sund}.

The starting point is Lagrangian $\tilde{L}$ as seen in \eqref{Lagn}.
Obviously there is no velocity for the degree of freedom $n$, which makes the
corresponding momenta $p_{n}=\frac{\partial\tilde{L}}{\partial\dot{n}}$ to be
zero and at the same time the Legendre transformation non-invertible. In such
cases the Dirac-Bergmann algorithm is used to pass to the Hamiltonian
description of the system. In our case $p_{n}\approx0$ is the primary
constraint of the theory, with ``$\approx$'' denoting a weak
equality\footnote{Having a quantity being weakly zero roughly signifies, that
it is not to be set to zero prior to any Poisson bracket calculation. Only the
end result is to be projected on the constrained surface where the constraints
vanish. For example, $p_{n}\approx0$ means that $\{n,p_{n}^{2}\}=2p_{n}=0$,
but $\{n,p_{n}\}=1$, i.e. the zero value of $p_{n}$ cannot be substituted when
the latter is inside a Poisson bracket.}. The primary constraint is added in
the Hamiltonian through a multiplier which is denoted here by $u_{n}$. The
$u_{n}$ can be seen as the missing velocity of $n$. Thus, the total
Hamiltonian is defined as
\begin{equation}
H_{T}=\frac{n}{2}\mathcal{H}+u_{n}p_{n} = \frac{n}{2}\left( G^{\alpha\beta
}p_{\alpha}p_{\beta}+\Lambda-\frac{3k}{a^{2}}\right)  + u_{n}p_{n}%
,\label{totalH}%
\end{equation}
where the momenta are given by $p_{\alpha}=\frac{\partial\tilde{L}%
}{\partial\dot{q}^{\alpha}}$ and $G^{\alpha\beta}$ is the inverse of the
minisuperspace metric \eqref{minimetric}. The $\mathcal{H}=G^{\alpha\beta
}p_{\alpha}p_{\beta}+\Lambda-\frac{3k}{a^{2}}$ is called the quadratic (or
Hamiltonian) constraint. It results as such due to the consistency requirement
that any of the constraints must be (at least through a weak equality)
preserved in time. Thus, for $p_{n}\approx0$, we need to have $\dot{p}_{n}
\approx0 \Rightarrow\{p_{n},H_{T}\} \approx0$, which leads to
\begin{equation}
\mathcal{H}=G^{\alpha\beta}p_{\alpha}p_{\beta}+\Lambda-\frac{3k}{a^{2}}%
\approx0. \label{qcon}%
\end{equation}
Hence, $\mathcal{H}\approx0$ becomes the secondary constraint of the
theory. No tetriary constraint is derived since the consistency condition
$\dot{\mathcal{H}}\approx0$ is satisfied identically. As a result our theory
has two constraints, namely $p_{n}\approx0$ and $\mathcal{H}\approx0$. They
additionally commute with each other $\{p_{n},\mathcal{H}\}=0$, which
categorizes them as first class constraints and signifies that the multiplier
$u_{n}$, as well as $n$, in \eqref{totalH} remain arbitrary functions of the
theory. The total Hamiltonian $H_{T}$ results in being a linear combination of
constraints.

Since at this level we deal with a simple quantum mechanics problem, we follow
the canonical quantization procedure. We thus choose the typical
representation where the positions act multiplicatively while the momenta are
first order linear operators (in what follows we work in $\hbar=1$ units)
\begin{equation}
q^{\alpha}\mapsto\widehat{q}^{\alpha}=q^{\alpha},\quad p_{\alpha}%
\mapsto\widehat{p}_{\alpha}=-\mathrm{i}\frac{\partial}{\partial q^{\alpha}}.
\end{equation}
We put in use Dirac's procedure for quantizing constrained systems and thus we
require that the constraints impose the following conditions upon the wave
function $\Psi$
\begin{align}
&  \widehat{p}_{n}\Psi=0\Rightarrow\frac{\partial\Psi}{\partial n}%
=0,\label{WDW}\\
&  \widehat{\mathcal{H}}\Psi=0.
\end{align}
The first relation simply states that $\Psi$ cannot depend on $n$ and the
second defines the well known Wheeler-DeWitt equation.

In order to address the factor ordering problem of the Hamiltonian constraint
operator, whose classical equivalent is \eqref{qcon}, we choose the conformal
Laplacian in order to express its kinetic term. The reason for this is twofold
\cite{tczan}: a) The wave functions that result as solutions of $\widehat
{\mathcal{H}} \Psi=0$ can give an invariant probability amplitude $dP=\mu
\Psi^{*} \Psi dV$ under transformations in the configuration space, where $dV$
is the corresponding volume element and $\mu=\sqrt{|G|}$, with $G=\mathrm{Det}%
(G_{\alpha\beta})$, the natural measure. b) The conformal Laplacian is
invariant under conformal transformations of the minisuperspace metric
\cite{ConfLap}. The latter is compatible with the scaling the lapse function
$N$ that we have used at the classical level. It is the freedom which we
exploited in order to pass from $N$ to $n$ and to the minisuperspace metric
\eqref{minimetric}. Hence we have
\begin{equation}
\label{Hop}\widehat{\mathcal{H}} = - \frac{1}{2\mu} \partial_{\alpha}(\mu
G^{\alpha\beta}\partial_{\beta}) + \frac{d-2}{8(d-1)} \mathcal{R} +
\Lambda-\frac{3 k}{a^{2}} ,
\end{equation}
where $\mathcal{R}$ is the Ricci scalar of the minisuperspace given in
\eqref{miniRicci} and $d=3$ its dimension.

The two additional operators that we need in order to distinguish a complete
set of states solving \eqref{WDW} are given by considering the quantization of
the $\mathfrak{so}(1,2)$ algebra. First we express the classical observables
$Q_{I}$ of \eqref{clasQ} as operators. To this end we adopt the most general
expression for a linear first order Hermitian operator under a measure $\mu$
\begin{equation}
\label{QIop}\widehat{Q}_{I}=-\frac{\mathrm{i}}{2\mu}\left(  \mu\xi_{I}%
^{\alpha}\partial_{\alpha}+\partial_{\alpha}\left(  \mu\xi^{\alpha}\right)
\right)  .
\end{equation}
The fact that we use the physical measure $\mu=\sqrt{|G|}$, together with the
$\xi_{I}$ being Killing vector fields of $G_{\alpha\beta}$, reduces the
generic expression \eqref{QIop} to just $\widehat{Q}_{I}=-\mathrm{i}\,\xi_{I}$
\cite{tchriSw}. In our case the vectors $\xi_{I}$ are those presented in
\eqref{sogen}. By construction, the $\widehat{Q}_{I}$ commute with the
Hamiltonian constraint operator of \eqref{Hop}.

In this setting the inner product between two states, which are
characterized by the wavefunctions $\Phi$ and $\Psi$, is given by
\begin{equation}
\label{innerpro}\langle\Phi| \Psi\rangle= \int\mu\Phi^{*} \Psi dV = \int
\sqrt{|G|} \Phi^{*} \Psi dV \propto\int a^{8} |\sinh\left( \lambda\phi\right)
| \Phi(a,\phi,\chi)^{*} \Psi(a,\phi,\chi) da\, d\phi\, d\chi,
\end{equation}
where we have substituted $G\propto-a^{16}\sinh^{2}(\lambda\phi)$ as
the determinant of the minisupermetric \eqref{minimetric}. There is an overall
numerical factor in the determinant which without loss of generality we can
ignore. When we later construct the orthonormal states such a numerical factor
can always be absorbed inside the normalization constant. Only the functional
dependence of $G$ on the configuration space variables $(a,\phi,\chi)$ is
effectively important. As we mentioned before, this choice of measure,
together with the operators we adopted, results in the inner product
$\langle\Phi| \Psi\rangle$ being invariant under diffeomorphisms of the
minisuperspace metric $G_{\alpha\beta}$ \cite{tczan}. We thus carry a
geometric property of the classical minisuperspace - the fact that the latter
is invariant under diffeomorphisms - to the quantum level of the inner product
and subsequently to the definition of the probability. In what follows below
we perform various transformations in the $(a,\phi,\chi)$ variables depending
on which offers in each case a better simplification of the results. With the
adoption of the square root of the determinant of the minisuperspace metric as
a measure we have guaranteed that this does not affect the quantization
process.

In order to proceed with the quantization, we need to use $\widehat{Q}_{3}$,
the Casimir invariant of the algebra $\widehat{K}=\widehat{Q}_{3}^{2}%
-\widehat{Q}_{1}^{2}-\widehat{Q}_{2}^{2}$ and of course $\widehat{\mathcal{H}%
}$ as a constraint on the wave function. We already know that the
$\mathfrak{so}(1,2)$ algebra - which is closely related to the
P\"{o}schl-Teller problem in quantum mechanics - results in states
characterized by two eigenvalues $m$, $\ell$ with
\begin{equation}
\widehat{Q}_{3}|m,\ell\rangle=m\lambda|m,\ell\rangle,\quad\widehat{K}%
|m,\ell\rangle=\ell(\ell+1)\lambda^{2}|m,\ell\rangle,
\end{equation}
where the constant factor $\lambda$ (the same appearing in the starting action
\eqref{action}) has been introduced at the right hand side for simplification
reasons. Alternatively - in what regards these two equations - we could also
absorb it inside $\phi$ and $\chi$.

The main difference with the typical angular momentum quantization of
$\mathfrak{so}(3)$ is that $\ell(\ell+1)$ may also assume negative values. The
significance of the $\mathfrak{so}(1,2)$ algebra in the quantum cosmology of
the axion - dilaton system, has already been observed in
\cite{Maharana1,Maharana2,Cordero}, where a specific coupling inspired by
string theory is assumed in the field kinetic terms in the context of a
positive spatial curvature FLRW spacetime without cosmological constant. The
$\mathfrak{so}(1,2)$ algebra quantization has also been connected with systems
obtained in a minisuperspace procedure \cite{Dimakis1,Kara}.

We start from the general case and present the solutions that we derive, but
restrain our in depth analysis for the maximal symmetry case of $\lambda=
\pm\sqrt{\frac{3}{2}}$, $k=0$ (as we also did at the classical level).
Nevertheless, a large part of what follows remains true for a generic
$\lambda$ as well.

More analytically, the three equations that need to be solved in order to
derive the wave function are $\widehat{Q}_{3}\Psi=m\lambda\Psi$, $\widehat
{K}\Psi=\ell(\ell+1)\lambda^{2}\Psi$ and $\widehat{\mathcal{H}}\Psi=0$, where
$m\lambda$ and $\ell(\ell+1)\lambda^{2}$ are the eigenvalues of $\widehat
{Q}_{3}$ and the Casimir operator $\widehat{K}$ respectively. The explicit
form of the three equations is:%
\begin{align}
\mathrm{i}\partial_{\chi}\Psi+m\lambda\Psi &  =0,\label{Q6ein1}\\
\partial_{\phi}^{2}\Psi+\lambda\coth(\lambda\phi)\partial_{\phi}\Psi+\frac
{1}{\sinh^{2}(\lambda\phi)}\partial_{\chi}^{2}\Psi-\ell(\ell+1)\lambda^{2}\Psi
&  =0,\label{Q6ein2}\\
\frac{1}{6a^{4}}\left(  \frac{1}{4}\partial_{a}^{2}\Psi+\frac{1}{a}%
\partial_{a}\Psi\right)  -\frac{1}{4a^{6}}\widehat{K}\Psi-\left(
\frac{2\lambda^{2}-3}{32a^{6}}+\frac{3k}{2a^{2}}-\frac{\Lambda}{2}\right)
\Psi &  =0. \label{Q6ein3}%
\end{align}

Due to the symmetry structure of the problem, the solution can be extracted by
splitting variables $\Psi=\psi_{1}(\chi)\psi_{2}(\phi)\psi_{3}(a)$.

The situation is quite simple in what regards \eqref{Q6ein1} since it implies
the solution
\begin{equation}
\psi_{1}(\chi)=\frac{1}{\sqrt{2\pi}}e^{\mathrm{i}m\lambda\chi},\quad
m\in\mathbb{Z} . \label{psi1}%
\end{equation}
The constant $m$ needs to be an integer due to the product $\lambda\chi$ being
a periodic variable, which leads us to impose the boundary condition:
$\psi_{1}(0)=\psi_{1}(2\pi/\lambda)$.

In what regards \eqref{Q6ein2}, after the splitting of variables, we obtain
\begin{equation}
\frac{1}{\sinh(\lambda\phi)}\frac{d}{d\phi}\left(  \sinh(\lambda\phi
)\frac{d\psi_{2}}{d\phi}\right)  -\lambda^{2}\left(  \ell(\ell+1)+\frac{m^{2}%
}{\sinh^{2}(\lambda\phi)}\right)  \psi_{2}=0, \label{casv}%
\end{equation}
which has the general solution
\begin{equation}
\psi_{2}(\phi)=C_{1}P_{\ell}^{m}(\cosh(\lambda\phi))+C_{2}Q_{\ell}^{m}%
(\cosh(\lambda\phi)), \label{psi2}%
\end{equation}
where $P_{\ell}^{m}$ and $Q_{\ell}^{m}$ are the associated Legendre functions
of the first and the second kind respectively, while $C_{1},~C_{2}$ are
integration constants.

The last equation to be addressed is \eqref{Q6ein3} which, by virtue of
\eqref{psi1} and \eqref{psi2}, becomes
\begin{equation}
4a\left(  a\frac{d^{2}\psi_{1}}{da^{2}}+4\frac{d\psi_{1}}{da}\right)
+3\left(  16a^{6}\Lambda-48a^{4}k-2(1+2l)^{2}\lambda^{2}+3\right)  \psi_{1}=0.
\label{wdwODE}%
\end{equation}
Linear ordinary differential equations of this form, involving polynomial
coefficients, are solved by holonomic functions. The latter can be defined by
the equation itself together with a set of boundary conditions. For specific
values of the parameter involved we can obtain well known functions; for example:

In the case $\Lambda=0$ the function $\psi_{1}(a)$ reads
\begin{equation}
\psi_{1}(a)=a^{-\frac{3}{2}}\left(  C_{3}I_{\frac{1}{2}\sqrt{\frac{3}{2}%
}\lambda(2l+1)}(3a^{2}\sqrt{k})+C_{4}I_{-\frac{1}{2}\sqrt{\frac{3}{2}}%
\lambda(2l+1)}(3a^{2}\sqrt{k})\right)  ,
\end{equation}
where $I_{\nu}(x)$ is the modified Bessel function of the first kind and
$C_{3},~C_{4}$ are integration constants.

When we consider a spatially flat universe, that is, $k=0$, we get
\begin{equation}
\psi_{1}(a)=a^{-\frac{3}{2}}\left(  C_{3}J_{\frac{\lambda(2l+1)}{\sqrt{6}}%
}(\frac{2a^{3}\sqrt{\Lambda}}{\sqrt{3}})+C_{4}J_{-\frac{\lambda(2l+1)}%
{\sqrt{6}}}(\frac{2a^{3}\sqrt{\Lambda}}{\sqrt{3}})\right)  , \label{psi1kzero}%
\end{equation}
where this time $J_{\nu}(x)$ is the Bessel function of the first kind and
$C_{3},~C_{4}$ again signify arbitrary integration constants.

As we previously mentioned we are interested to make a study in the special
case $k=0$, $\lambda= \pm\sqrt{\frac{3}{2}}$ which exhibits the highest level
of symmetry.

\section{Quantum analysis for $k=0$, $\lambda=\pm\sqrt{\frac{3}{2}}$}

\label{sec5}

We noticed that the eigenvalue $\ell(\ell+1)$ can assume negative and
non-negative values. We shall refer to both cases separately in what follows.
But first it is useful to stress that, on mass shell, the value that the
classical counterparts of $\widehat{Q}_{3}$ and $\widehat{K}$ are%
\begin{equation}
Q_{3}=\sqrt{2}\frac{a_{0}^{3}\gamma}{\alpha}~, \label{kclas1}%
\end{equation}%
\begin{equation}
K=Q_{3}^{2}-Q_{1}^{2}-Q_{2}^{2}=\frac{2a_{0}^{6}(1-\alpha^{2})\gamma^{2}%
}{\alpha^{2}}. \label{Kclas}%
\end{equation}
Again we note that the arithmetic value of $a_{0}$ is irrelevant. At
the classical level, and given that $k=0$, it can be normalized to $|a_{0}|=1$
with a diffeomorphism in $r$, which is the radial variable in line element
\eqref{FLRW}. We need only use it appropriately so as to keep the expression
\eqref{asol1} for $a(t)$ real when necessary.

\subsection{The $\ell(\ell+1)\geq0$ case}

By looking at \eqref{Kclas}, we conclude that positive values for $K$
correspond to the phantom epoch, $w<-1$, of our classical analysis. The latter
can be reproduced under two conditions: (A) By assuming both $\alpha$ and
$\gamma$ to be imaginary (then, as we saw in the classical bouncing solution,
$a_{0}$ is normalized to $a_{0}=1$ or $a_{0}=e^{-\mathrm{i} \frac{\pi}{3}}$)
or (B) by considering only $\gamma$ to be imaginary, but $-1<\alpha<1$ (which
requires $a_{0}=e^{\pm\mathrm{i}\frac{\pi}{6}}$). We separately study these
two possibilities that belong to the same class of having $\ell(\ell+1)\geq0$.

\begin{enumerate}
[(A)]

\item We start with the first case, which can be seen as a direct analogy to
what happens in the typical angular momentum quantization. Truly, the
consequence of $\alpha$ being imaginary is that the expression for $\phi$ also
becomes imaginary. If we take this fact in account at the quantum level by
introducing the variable $v=-\mathrm{i}\lambda\phi$, $v\in\mathbb{R}$ in
\eqref{psi2}, the first branch of the solution is written as
\begin{equation}
\psi_{2}(v)=C_{1}P_{\ell}^{m}(\cos v)~,
\end{equation}
which is what one obtains from the $\mathfrak{so}(3)$ quantization. This is a
normal consequence of the fact that at the classical level, the change
$\phi=\frac{\mathrm{i}}{\lambda}v$ signifies that the minisuperspace metric
\eqref{minimetric} becomes Euclidean (with an overall minus sign). Thus, if we
consider the properties of the classical solution the part of the wave
function involving the $\phi$ (or the $v$ under our substitution) dependence
is given by $P_{\ell}^{m}(\cos v)$ which is well known that satisfies the
orthogonality relation
\begin{equation}
\int_{-1}^{1}P_{\ell}^{m}(\cos v)P_{\ell}^{m}(\cos v)d(\cos v)=\frac{2}%
{2\ell+1}\frac{(\ell+m)!}{(\ell-m)!}\delta_{\ell k},\quad|m|\leq\ell,\quad
\ell\in\mathbb{N}.
\end{equation}

Note that the natural measure adopted in \eqref{innerpro} is what is
needed here in order to obtain the correct weight function in the above
orthogonality relation. If we consider that $v=-\mathrm{i} \lambda\phi$, we
obtain
\begin{equation}
\sinh(\lambda\phi) d\phi\propto- \sin(v) dv = d (\cos v) .
\end{equation}
The overall constant factor resulting from the transformation is of
course absorbable in the normalization constant.

\item In the second case, where only $\gamma$ is imaginary, the classical
behaviour of the system implies that $\lambda\phi$ has a constant imaginary
part, $\pm\mathrm{i}\frac{\pi}{2}$, depending on whether $\mathrm{Re}%
(\lambda\phi)$ is negative or positive.

In the analysis that follows the sign of the constant imaginary part plays no
role, so we choose only one sign and perform the transformation $\lambda
\phi=\tilde{v}-\mathrm{i}\frac{\pi}{2}$ in equation \eqref{casv} together with
a reparametrization of the function $\psi_{2}(\tilde{v})=\frac{f(\tilde{v}%
)}{\cosh^{1/2}\tilde{v}}$. This turns equation \eqref{casv} into
\begin{equation}
\label{pote1}\frac{d^{2}f}{d\tilde{v}^{2}}+\left[  \frac{m^{2}-\frac{1}{4}%
}{\cosh^{2}\tilde{v}}-\frac{(2\ell+1)^{2}}{4}\right]  f=0,
\end{equation}
where we recognize a special case of the general hyperbolic P\"{o}schl -
Teller potential \cite{Poschl}. A complete analysis can be found in
\cite{Landau,Poschl2} but, for the sake of completeness, we provide a brief
description of the problem. In the case of a particle of mass $m_{0}$ moving
under the influence of the P\"{o}schl - Teller potential, the one dimensional,
time independent Schr\"odinger equation reads
\begin{equation}
\label{pote2}\frac{d^{2}\tilde{\Psi}}{dx^{2}}+ \frac{2m_{0}}{\hbar^{2}}\left[
E+ \frac{U_{0}}{\cosh^{2}(\alpha_{0} x)}\right]  \tilde{\Psi} =0,
\end{equation}
where $E$ is the energy of the particle and $\alpha_{0}$, $U_{0}$
constants characterizing the breadth and the depth of the potential $U(x)=-
\frac{U_{0}}{\cosh^{2}(\alpha_{0} x)}$ respectively. In Fig. \ref{Fig70} we
give a graphic representation of the potential. For positive values of the
energy the spectrum is continuous, while for negative it becomes discrete
since the particles's motion is now constrained by the potential. What is
more, the number of the energy levels is bounded. In particular it can be
shown that the bounded spectrum is given by
\begin{equation}
E_{n} = -\frac{\hbar^{2} \alpha_{0}^{2}}{8 m_{0}} \left[ -(1+2 n) + \sqrt{1+
\frac{8m_{0} U_{0}}{\alpha_{0}^{2} \hbar^{2}}} \right] ,
\end{equation}
where $n\in\mathbb{N}$ and at the same time $n=\beta-\epsilon$, with
\begin{equation}
\epsilon^{2} = - \frac{2m_{0} E}{\hbar^{2} \alpha_{0}^{2}} \quad\text{and}
\quad p (p+1) = \frac{2m_{0} U_{0}}{\alpha_{0}^{2} \hbar^{2}} .
\end{equation}

\begin{figure}[ptb]
\centering\textbf{ \includegraphics[width=0.5\textwidth]{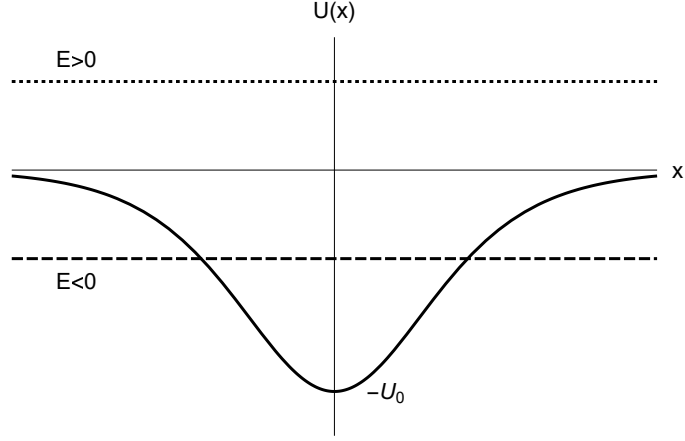}
}\caption{The P\"{o}schl - Teller potential. If the particle has positive
energy $E>0$ its motion is unbounded, else ($E<0$) the spectrum of the energy
becomes discrete.}%
\label{Fig70}%
\end{figure}

By comparing \eqref{pote1} and \eqref{pote2}, the analogy is obvious. The
constants $\alpha_{0}$, $\hbar$ are all set to unity, while $m_{0}=\frac{1}%
{2}$ and the ``energy'' of the particle of our case becomes $E=-\epsilon
^{2}=-\frac{(2\ell+1)^{2}}{4}$. It is negative and thus it results in a finite
number of bound states. Additionally we have $p(p+1)=m^{2}-\frac{1}{4}$ and
the eigenvalues follow the restriction
\begin{equation}
p-\epsilon=n,\quad n\in\mathbb{N},\quad n<p.
\end{equation}

The wave function solving \eqref{pote1} is given in terms of the
associated Legendre polynomials $f(\tilde{v}) \propto(-1)^{p-n} P^{p-n}%
_{p}(\tanh(\tilde{v}))$ \cite{Poschl3,SDong}, which satisfy the normalization
condition
\begin{equation}
\label{assPorth}\int_{0}^{1} \frac{\left[ P^{q}_{p}(s)\right] ^{2}}{1-s^{2}}
ds = \frac{1}{2 q} \frac{\left( p+q\right) !}{\left( p-q\right) !}, \quad0 <
q<p .
\end{equation}
The weight factor inside the integral of \eqref{assPorth} is exactly
what we obtain from the measure choice we did in \eqref{innerpro}. If we
concentrate on the $\phi$ dependent part we have
\begin{equation}
\sinh(\lambda\phi) |\psi_{2}(\phi)|^{2} d\phi\propto|f(\tilde{v})|^{2}
d\tilde{v} \propto\left[ P^{p-n}_{p}(\tanh(\tilde{v}))\right] ^{2} d\tilde{v}
= \frac{\left[ P^{p-n}_{p}(s)\right] ^{2}}{1-s^{2}} ds,
\end{equation}
where in the above the subsequent changes $\lambda\phi= \tilde{v} -
\mathrm{i} \frac{\pi}{2}$, $\psi_{2}(\tilde{v})=\frac{f(\tilde{v})}%
{\cosh^{1/2}\tilde{v}}$ and $\tanh\tilde{v}=s$ have been used.
\end{enumerate}

In what regards both of the previous cases, the dependence of the wave
function from the scale factor $a$ is decided from the remaining
Wheeler-DeWitt equation \eqref{Q6ein3} and its solution \eqref{psi1kzero},
which for $\lambda=\pm\sqrt{\frac{3}{2}}$ and under a change of variables
$a=\left(  \frac{3u^{2}}{4}\right) ^{1/6}$, it can be written as
\begin{equation}
\psi_{3}(u)=\widetilde{C}_{3}j_{\ell}(\pm\sqrt{\Lambda}u)+\widetilde{C}%
_{4}y_{\ell}(\pm\sqrt{\Lambda}u), \label{sphbess}%
\end{equation}
where $\widetilde{C}_{3},~$and $\widetilde{C}_{4}$ are linear combinations of
the integration constants $C_{3}$ and $C_{4}$ of (\ref{psi1kzero})
respectively. We remind here the basic formulas connecting the Bessel
functions, $Y_{\nu}(z) = \frac{J_{\nu}(z)\cos(\nu\pi) - J_{-\nu}(z)}{\sin
(\nu\pi)}$, $j_{\nu}(z) = \sqrt{\frac{\pi}{2z}} J_{\nu+\frac{1}{2}}(z)$,
$y_{\nu}(z) = \sqrt{\frac{\pi}{2z}} Y_{\nu+\frac{1}{2}}(z)$, for $\nu,z$ being
generally complex variables.

For the wave function we choose the first branch of the solution, since it is
well known that the spherical Bessel function of the first kind satisfies the
orthogonality relation
\begin{equation}
\int_{0}^{+\infty}u^{2}j_{\ell}(\sigma u)j_{\ell}(\sigma^{\prime}%
u)du=\frac{\pi}{2\sigma\sigma^{\prime}}\delta(\sigma-\sigma^{\prime}).
\label{SBorth}%
\end{equation}
In our case $\sigma=\sigma^{\prime}=\pm\sqrt{\Lambda}$ which is allowed a
continuous set of values. Notice that again the natural measure of
\eqref{innerpro} is the one resulting in the needed weight in the integral
\eqref{SBorth} since, by concentrating on the $a=\left( \frac{3u^{2}}%
{4}\right) ^{1/6}$ dependence, we have
\begin{equation}
a^{8} da \propto u^{2} du .
\end{equation}
Once more the overall multiplicative numerical factor is irrelevant
since it can be absorbed in the normalization constant.

In order to sum the above results we write final wavefunctions. In the
first case where $\phi= \frac{\mathrm{i}}{\lambda} v$ we have
\begin{equation}
\Psi(u,v,\chi) = C_{A} e^{\mathrm{i} m \lambda\chi}\, P_{\ell}^{m}(\cos v)\,
j_{\ell}(\sqrt{\Lambda} u),
\end{equation}
where the quantum numbers follow the restrictions
\begin{equation}
m \in\mathbb{Z}, \quad\ell\in\mathbb{N}, \quad|m|\leq\ell
\end{equation}
and the normalization constant is
\begin{equation}
C_{A} = \frac{1}{\pi} \left[ \Lambda\left( \ell+\frac{1}{2}\right)
\frac{\left( \ell-m\right) !}{\left( \ell+m\right) !} \right] ^{\frac{1}{2}} .
\end{equation}
The measure used is $\mu_{A} dV= u^{2}\, du\, d\left( \cos(v)\right)
\,d\chi$ and it differs from what we introduced in \eqref{innerpro} by a
multiplicative constant. We remind here that the variable $u$ is related to
$a$ through $a=\left( \frac{3u^{2}}{4}\right) ^{1/6}$. For simplicity we use
directly $\mu_{A}$ in order to have a simpler expression for $C_{A}$, i.e. not
having to include in it an additional numerical factor.\footnote{Another
reason for not using directly the $\mu$ of \eqref{innerpro} is because the
latter was written given a real variable $\phi$. In the two cases considered
in this section $\phi$ is taken to be complex, thus we could not have adopted
exactly the same expression. Nevertheless the difference is just an overall
constant.}

The second case in which $\phi=\frac{\tilde{v}}{\lambda}-\mathrm{i}
\frac{\pi}{2\lambda}$ leads to a wave function
\begin{equation}
\Psi(u,\tilde{v},\chi) = C_{B} e^{\mathrm{i} m \lambda\chi}\, \frac
{P^{p-n}_{p}(\tanh(\tilde{v}))}{\left[ \cosh(\tilde{v})\right] ^{1/2}}\,
j_{\ell}(\sqrt{\Lambda} u),
\end{equation}
where the following relations hold
\begin{equation}
\label{evalue2}\epsilon^{2}=\frac{(2\ell+1)^{2}}{4}, \quad p(p+1)=m^{2}%
-\frac{1}{4}, \quad p-\epsilon=n \in\mathbb{N}, \quad m \in\mathbb{Z}%
\end{equation}
and the normalization constant is
\begin{equation}
C_{B} = \frac{(-1)^{p-n}}{\pi} \left[ 2 \Lambda\left( p-n\right)  \frac
{n!}{\left( 2p-n\right) !} \right] ^{\frac{1}{2}} .
\end{equation}
Notice that due to the \eqref{evalue2}, the $2p-n$ in the above
relation is also a natural number. For the measure in this case we have
$\mu_{B} dV = u^{2} \cosh(\tilde{v}) \, du \, d\tilde{v}\, d\chi$ and as
previously the difference with the $\mu$ of \eqref{innerpro} is once more a
multiplicative constant.

We have thus demonstrated, that the classical $w<-1$ situation, which is
obtained by two distinct choices of parameters, corresponds at the quantum
level to \textquotedblleft bound states". That is, a discrete spectrum for the
quantum numbers $m$ and $\ell$, although the quantum procedure in the two
cases is different (as also the quantum conditions on $\ell$). Additionally,
in our formulation of the system, the cosmological constant $\Lambda$ can be
perceived as a quantum eigenvalue from the Wheeler-DeWitt equation, which
however has a continuous spectrum and it is normalized to the Dirac delta.

Finally it is interesting to remark that the $\ell(\ell+1)=0$ eigenvalue has
its classical equivalent on having $\alpha=1$ which corresponds to the de
Sitter universe (remember that $\ell(\ell+1)$ is the eigenvalue of
$\widehat{K}$ whose classical counterpart's on mass shell value is given in \eqref{Kclas}).

\subsection{The $\ell(\ell+1)<0$ case}

From the classical solution we know that \eqref{Kclas} is negative whenever
both $\alpha$ and $\gamma$ are real and we saw that this corresponds to what
we characterized as the quintessence epoch. In this case, we can parameterize
the quantum number $\ell$ as $\ell=\mathrm{i}s-\frac{1}{2}$, $s\in\mathbb{R}$.
Then, the eigenvalue $\ell(\ell+1)$ remains real but it is exclusively
negative. We have of course the same general solution \eqref{psi2} of the
eigenvalue equation \eqref{casv}, only that now $\lambda\phi=\bar{v}$ is real
and we have to take into account the orthogonality relation \cite{Van}
\begin{equation}
\int_{1}^{+\infty}P_{\mathrm{i}s-1/2}^{m}(\cosh\bar{v})P_{\mathrm{i}s^{\prime
}-1/2}^{m}(\cosh\bar{v})d(\cosh\bar{v})=\frac{(-1)^{m}\coth(\pi s)}{s}%
\frac{\Gamma(\mathrm{i}s+\frac{1}{2}+m)}{\Gamma(\mathrm{i}s+\frac{1}{2}%
-m)}\delta(s-s^{\prime}), \label{Porth2}%
\end{equation}
where $\Gamma(x)$ denotes the gamma functions that extend the factorial definition.

The spectrum of $\ell(\ell+1)$ is continuous and the system corresponds to
\textquotedblleft free" states. In what regards \eqref{Porth2}, we have to
notice that in the literature there appear two different definitions of the
associated Legendre function. Under the first definition the function
$P_{\mathrm{i}s-1/2}^{m}(x)$ is real in the region $x>1$, which is our domain
of integration. The second definition differs from the first by a
multiplicative constant factor of $e^{\mathrm{i}\pi m/2}$. However, when the
product $(P^{m}_{\mathrm{i}s-1/2})^{\ast}P^{m}_{\mathrm{i}s-1/2}$ is used,
such a constant phase is eliminated. So, as far as the probability is
concerned, there is no distinction between the two definitions. For simplicity
and to avoid complex conjugates we use here the one according to which
$P_{\mathrm{i}s-1/2}^{m}(x)$ is a real function.

The only thing that remains is to decide which linear combination of the
solution of \eqref{sphbess} is appropriate to serve in our description. In
this case the spherical Bessel is of complex order $\ell=\mathrm{i}s-\frac
{1}{2}$. We may use the transformation $\psi_{3}(u)=\frac{1}{\sqrt{u}}%
\tilde{\psi}_{3}(u)$ in \eqref{sphbess} and derive the solution
\begin{equation}
\tilde{\psi}_{3}=\bar{C}_{3}J_{\mathrm{i}s}(\sqrt{\Lambda}u)+\bar{C}%
_{4}Y_{\mathrm{i}s}(\sqrt{\Lambda}u), \label{bessim}%
\end{equation}
where $\bar{C}_{3}$, $\bar{C}_{4}$ are constants of integration proportional
to the $\widetilde{C}_{3}$, $\widetilde{C}_{4}$ of \eqref{sphbess}
respectively. In the case of imaginary order an interesting linear combination
of the solution can be distinguished in the form of the function
\cite{Dunster}
\begin{equation}
\tilde{\psi}_{3}=C\tilde{J}_{s}(\sqrt{\Lambda}u)=\frac{C}{\cosh\left(  \pi
s/2\right)  }\mathrm{Re}[J_{\mathrm{i}s}(\sqrt{\Lambda}u)],
\end{equation}
where $C$ is related to linear combinations of the previous constants,
$\bar{C}_{3}$ and $\bar{C}_{4}$, and which serves to normalize the wave
function. By definition the $\tilde{J}_{s}(\sqrt{\Lambda}u)$ is a real
function. What is more, a normalization condition in terms of a Dirac delta
can also be derived (see appendix \ref{app1}).

Interestingly enough, the states spanned by different values of $\Lambda$ are
not necessarily orthogonal. It can be deducted however (see once more appendix
\ref{app1}) that the orthogonality condition between two states characterized
by eigenvalues $\Lambda$ and $\Lambda^{\prime}$ requires either $s=0$ or
\begin{equation}
\label{conLam}\frac{\Lambda}{\Lambda^{\prime}}=e^{\frac{2k^{\prime}\pi}{s}%
},\quad k^{\prime}\in\mathbb{Z}.
\end{equation}
The requirement emerges from the need to cancel the lower limit of the
integral when the inner product is used. However, it has been shown that if a
different linear combination of the solution is adopted, such a contribution
can be avoided, see \cite{Gryb,Gielen}. According to the choice of the
solution we make here the $u$ dependence of the wave function is (remember
that $u$ is the variable related to the scale factor $a$)
\begin{equation}
\psi_{3}(u)\propto\frac{1}{\sqrt{u}}\tilde{J}_{s}(\sqrt{\Lambda}u),
\end{equation}
and thus the full wave function becomes
\begin{equation}
\Psi(u,\bar{v},\chi) = C e^{\mathrm{i} m \lambda\chi} P_{\mathrm{i}s-1/2}%
^{m}(\cosh\bar{v}) \frac{1}{\sqrt{u}}\tilde{J}_{s}(\sqrt{\Lambda}u),
\end{equation}
where $m \in\mathbb{Z}$, $s\in\mathbb{R}$ and $\Lambda$ is restricted
by \eqref{conLam}, while the normalization constant reads
\begin{equation}
\label{NC3}C = \left[ \frac{s \sqrt{\Lambda}\; \Gamma(\mathrm{i} s+\frac{1}%
{2}-m)}{2 \pi(-1)^{m} \coth(\pi s) \Gamma(\mathrm{i} s+\frac{1}{2}+m)} \right]
^{\frac{1}{2}},
\end{equation}
under the inner product of $u^{2} \, du^{2} \, d\left(  \cosh(\bar{v})
\right)  d\chi$, where $\bar{v}=\frac{\phi}{\lambda}$.

In contrast to what we found in the phantom epoch, $w<-1$, which is described
by a bounded set of states, the quintessence era is characterized by a
continuous spectrum for $\widehat{K}$. We have to note that the eigenvalue $m$
remains discrete in both cases since the sign of $\ell(\ell+1)$ makes no
difference for it.

Lastly we want to address what happens at the classical singularity,
which is present in the quintessence epoch geometry. We recognize that there
is no general consensus on what constitutes a singularity avoidance at the
quantum level, since there exist various different methods of quantization and
in many cases even different interpretations of the wave function and of the
corresponding probability within the same theory. However, a reasonable
assumption is to not have divergences in the region in which a classically
problematic point corresponds \cite{Kiefer}. In our case this happens when
$a\rightarrow0\Rightarrow u\rightarrow0$ when we consider the quintessence
case. From our choice of inner product we can see that the probability
amplitude for the $u$ dependent part of the wave function is analogous to
\begin{equation}
\label{singavoid}\rho_{u}=u^{2}\psi_{3}(u)^{\ast}\psi_{3}(u) =
\begin{cases}
u^{2} j_{\ell}(\sqrt{\Lambda}u) j_{\ell}(\sqrt{\Lambda}u), & \mbox{if}
\quad\ell(\ell+1)\geq0\\
u \tilde{J}_{s}(\sqrt{\Lambda}u) \tilde{J}_{s}(\sqrt{\Lambda}u), & \mbox{if}
\quad\ell(\ell+1)<0 .
\end{cases}
\end{equation}
We can see that in both cases the limit $u\rightarrow0$ leads to
$\rho_{u}\rightarrow0$. In the first branch of $\ell(\ell+1)\geq0$,
corresponding classically to the phantom solution, we have at the small $u$
limit: $j_{\ell}(\sqrt{\Lambda}u) \sim u^{\ell}$. So, $\rho_{u}$ clearly goes
to zero as $u\rightarrow0$, even though classically we have no particular
problem at the given point. For the second branch, which at the classical
level relates to the quintessence solution, the behaviour of the function
$\tilde{J}_{s}(\sqrt{\Lambda}u)$ for small arguments can be found in
\eqref{asymJ} of appendix \ref{app1}. It oscillates strongly but it remains
finite. Thus the $u$ factor in the second part of \eqref{singavoid} dominates
sending $\rho_{u}$ to zero as we approach $u=0$. We interpret this result as a
zero transition probability to the problematic point $u=0$ from a neighboring
point $u\neq0$, which we consider to be a good sign for avoiding the
singularity.

\section{Conclusions}

\label{sec6}

In this work we consider a two-scalar field cosmology. Specifically, we assume
the contribution in the Einstein field equations only by the kinetic parts of
two interacting scalar fields as in the $\alpha$-attractor model, where the
fields are also minimally coupled to gravity. In addition, we assume that the
scalar field potential is constant and thus plays the role of a cosmological
constant. In a FLRW background space we derived the generic classical analytic
solution of this model.

In the case of a spatially flat universe and for a specific value of one of
the free parameters of the model, we were able to write the analytic solution
in closed form. For that specific case we studied the behaviour of the
physical parameters for various ranges of the integration constants of the
problem. Surprisingly, we found that it is possible to describe a quintessence
epoch, a phantom epoch or even a quintom model depending on the values of the
free parameters. In all of the above configurations appropriate domains of
deinition for the parameters can be found so that the physical quantities are
real functions. The quintessence solution describes a universe with an initial
singularity, in contrast to the phantom (or the quintom) solution which
results in a bouncing universe. In addition, all of the solutions have a
common future: at late times they give the de Sitter universe as an attractor.

Successively, we applied the minisuperspace approach to perform a canonical
quantization and write the Wheeler-DeWitt equation of quantum cosmology. We
showed how to construct quantum operators from the classical conservation
laws, which are used as additional constraints in order to calculate the wave
function of the universe. We perform a preliminary analysis for the generic
case and we focus the construction of a valid set of states for the particular
configuration which we investigated at the classical level and which
corresponds to the spatially flat FLRW universe. We demonstrate how different
eigenvalues of the quantum operators are related with the two different
behaviours of the classical solutions, that is, different quantum states
describe distinct classical behaviours.

It is very interesting the fact that an expansion characterized by a phantom
behaviour, i.e. $w<-1,$ is related to a bounded set of states, while on the
other hand the quintessence epoch, that is;$\ w>-1,$ corresponds to free
states for the universe. In this sense we see a hint of realization of how
canonical quantization of gravity, as initially introduced in the seminal
paper by B. S. DeWitt \cite{DeWitt}, tries to take form as a quantization in
the space of geometries, with different sets of states being linked to
different classical geometries. Another important remark is that the resulting
wave function leads to a resolution of the classical singularity (when the
latter is present) in the sense that the probability amplitude tends to zero
at the problematic point.

\appendix

\section{}

\label{app0}

Here we give the details on how to arrive from \eqref{akzero} to
\eqref{asol1}, when $\lambda=\pm\sqrt{\frac{3}{2}}$ and the gauge choice
$\chi(t)=\pm\sqrt{\frac{2}{3}} t$ has been made. Solution \eqref{akzero} is
derived for $k=0$ and when we set the aforementioned values for $\lambda$ and
$\chi$ it reads
\begin{equation}
a(t) = \left[ \frac{c_{2}}{\sinh\left[ \tanh^{-1}\left( \sqrt{\alpha^{2}%
-1}\tan\left( t+\beta\right) \right) +\bar{c}_{1}\right] } \right] ^{\frac
{1}{3}} .
\end{equation}
First, we reparametrize the constant $\bar{c}_{1}$ as $\bar{c}_{1} =
\ln\tilde{c}_{1}$ and exploit the identity
\begin{equation}
\tan^{-1} \left( \sqrt{\alpha^{2}-1}\tan\left( t+\beta\right) \right)  =
\frac{1}{2}\ln\left(  1+ \sqrt{\alpha^{2}-1} \tan\left( t+\beta\right)
\right)  -\frac{1}{2}\ln\left(  1- \sqrt{\alpha^{2}-1} \tan\left(
t+\beta\right) \right) ,
\end{equation}
in order to write
\begin{equation}
\label{intsnh}%
\begin{split}
& \sinh\left[ \tanh^{-1}\left( \sqrt{\alpha^{2}-1}\tan\left( t+\beta\right)
\right)  +\bar{c}_{1}\right]  = \frac{-1+ \sqrt{\alpha^{2}-1}\tan\left(
t+\beta\right)  + \tilde{c}_{1}^{2} \left( 1+\sqrt{\alpha^{2}-1}\tan\left(
t+\beta\right) \right) }{2\tilde{c}_{1} \sqrt{1+ \left( 1-\alpha^{2}\right)
\tan^{2}\left( t+\beta\right) }}\\
&  = \frac{1}{\sqrt{1-\alpha^{2} \sin^{2}\left( t+\beta\right) }} \left[
\frac{\tilde{c}_{1}^{2} -1}{2\tilde{c}_{1}} \cos\left( t+\beta\right)  +
\frac{1+\tilde{c}_{1}^{2}}{2\tilde{c}_{1}}\sqrt{\alpha^{2}-1} \sin\left(
t+\beta\right) \right] .
\end{split}
\end{equation}
With a subsequent reparametrization of the constant $c_{2}$ as
$c_{2}=- \mathrm{i} \frac{a_{0}^{3}}{\alpha\tilde{c}_{2}}$, we can use
\eqref{intsnh} to express the cube of the scale factor, $a(t)^{3}$, as
\begin{equation}
\label{cuba}a(t)^{3} = \frac{\frac{a_{0}^{3}}{\alpha\tilde{c}_{2}}
\sqrt{\alpha^{2} \sin^{2} \left( t+\beta\right) -1}}{\frac{\tilde{c}_{1}^{2}
-1}{2\tilde{c}_{1}} \cos\left( t+\beta\right)  + \frac{1+\tilde{c}_{1}^{2}%
}{2\tilde{c}_{1}}\sqrt{\alpha^{2}-1} \sin\left( t+\beta\right) } .
\end{equation}
At this point note that in place of $c_{2}$ we introduced two
constants, $\tilde{c}_{2}$ and $a_{0}$. We are going to fix appropriately the
former, while the latter remains arbitrary. We proceed by reparametrizing
$\tilde{c}_{1}$ and fixing $\tilde{c}_{2}$ in such a manner so that we have
\begin{equation}
\tilde{c}_{2} \frac{\tilde{c}_{1}^{2} -1}{2\tilde{c}_{1}} = \cos\left(
\zeta+\beta\right) \quad\text{and} \quad\tilde{c}_{2} \frac{1+\tilde{c}%
_{1}^{2}}{2\tilde{c}_{1}}\sqrt{\alpha^{2}-1} = \sin\left( \zeta+\beta\right) ,
\end{equation}
where $\zeta$ is the newly introduced constant in place of $\tilde
{c}_{1}$. As a result \eqref{cuba} becomes
\begin{equation}
\label{cuba2}a(t)^{3} = a_{0}^{3} \frac{\sin\left( t+\beta\right) }{\cos\left(
t- \zeta\right) } \sqrt{1- \frac{1}{\alpha^{2} \sin^{2} \left( t+\beta\right)
}} .
\end{equation}

We can exploit the invariance of the solution under time translations
to eliminate one of the two additive constants, either $\beta$ or $\zeta$. For
example, we can perform a time translation $t\mapsto t+\zeta$ and at the same
time introduce a new constant $B=\beta+\zeta$. Then only $B$ enters the
expressions for $a$, $N$ and $\phi$, while $\zeta$ just appears additively in
$\chi$. By demanding the boundary condition $\chi\left(  0\right)  =0$ we can
set $\zeta$ to be zero. We avoid all this procedure and straightforwardly set
$\zeta=0$ in \eqref{cuba2}. By successively taking the cubic root of
\eqref{cuba2}, we are finally led to equation \eqref{asol1}.

\section{}

\label{app1}

Let us start from the Bessel equation of our case which is satisfied by the
solution \eqref{bessim}
\begin{equation}
\frac{d}{du}\left(  u\frac{dW_{s,\sigma}}{du}\right)  +\left(  \frac{s^{2}}%
{u}+\sigma^{2}u\right)  W_{s,\sigma}=0,
\end{equation}
where as an eigenvalue we have set $\sigma=\pm\sqrt{\Lambda}$ and with
$W_{s,\sigma}$ we denote the solution that we use, the real function
\begin{equation}
W_{s,\sigma}=\tilde{\psi}_{3}(u)=\tilde{J}_{s}(\sigma u)=\frac{1}{\cosh\left(
\pi s/2\right)  }\mathrm{Re}[J_{\mathrm{i}s}(\sigma u)].
\end{equation}

We follow the usual procedure with which we derive orthogonality conditions in
a Sturm - Liouville problem: We take the equation for a different eigenvalue
$\sigma^{\prime}=\pm\sqrt{\Lambda^{\prime}}\neq\pm\sqrt{\Lambda}=\sigma$
\begin{equation}
\frac{d}{du}\left(  u\frac{dW_{s,\sigma^{\prime}}}{du}\right)  +\left(
\frac{s^{2}}{u}+\sigma^{\prime2}u\right)  W_{s,\sigma^{\prime}}=0.
\end{equation}
We multiply the first equation with $W_{s,\sigma^{\prime}}$ and the second
with $W_{s,\sigma}$ and subsequently we subtract by parts. As a result we
arrive at
\begin{equation}
(\sigma^{2}-\sigma^{\prime2})uW_{s,\sigma}W_{s,\sigma^{\prime}}=\frac{d}%
{du}\left[  u\left(  W_{s,\sigma}\frac{d}{du}W_{s,\sigma^{\prime}}%
-W_{s,\sigma^{\prime}}\frac{d}{du}W_{s,\sigma}\right)  \right]
\end{equation}
from where integration over the half line $\mathbb{R}_{+}$ leads to
\begin{equation}
(\sigma^{2}-\sigma^{\prime2})\int_{0}^{+\infty}\!\!uW_{s,\sigma}%
W_{s,\sigma^{\prime}}du=\left[  \mathcal{A}\right]  _{0}^{+\infty}:=\left[
u\left(  W_{s,\sigma}\frac{d}{du}W_{s,\sigma^{\prime}}-W_{s,\sigma^{\prime}%
}\frac{d}{du}W_{s,\sigma}\right)  \right]  _{0}^{+\infty}. \label{orthapp2}%
\end{equation}

If we take into account the approximate expressions for $\tilde{J}_{s}(x)$
\begin{equation}
\label{asymJ}\tilde{J}_{s}(x)\sim%
\begin{cases}
\sqrt{\frac{2}{\pi x}}\cos(x-\frac{\pi}{4}), & \mbox{if}\;x\rightarrow
+\infty\\
\left(  \frac{2\tanh\left(  \frac{\pi s}{2}\right)  }{\pi s}\right)
^{\frac{1}{2}}\cos\left(  s\ln\left(  \frac{x}{2}\right)  -\gamma_{s}\right)
, & \mbox{if}\;x\rightarrow0^{+},
\end{cases}
\end{equation}
where $\gamma_{s}$ is a specific constant depending on $s$, then it is easy to
see that
\begin{equation}
\mathcal{A}_{0}:=\lim_{u\rightarrow0}\mathcal{A}=\frac{2}{\pi}\tanh\left(
\frac{\pi s}{2}\right)  \sin\left[  s\ln\left(  \frac{\sigma}{\sigma^{\prime}%
}\right)  \right]  \label{lim0app2}%
\end{equation}
and
\begin{equation}
\mathcal{A}_{\infty}:=\lim_{u\rightarrow+\infty}\mathcal{A}=\frac{1}{\pi
\sqrt{\sigma\sigma^{\prime}}}\Big[(\sigma+\sigma^{\prime})\sin\left[
(\sigma-\sigma^{\prime})u\right]  -\left(  \sigma^{\prime}-\sigma\right)
\cos\left[  (\sigma+\sigma^{\prime})u\right]  \Big]. \label{liminfapp2}%
\end{equation}
This last term gives a delta function when interpreted in a distributional
sense, as we are going to see below. However, we may notice that the other
contribution in \eqref{orthapp2}, given by \eqref{lim0app2} is not zero. This
means that the states with $\sigma\neq\sigma^{\prime}$ are not necessarily orthogonal.

Only for $s=0$ or for those eigenvalues $\sigma,\sigma^{\prime}$ that satisfy
\begin{equation}
\ln\left(  \frac{\sigma}{\sigma^{\prime}}\right)  =\frac{k^{\prime}\pi}%
{s}\Rightarrow\ln\left(  \frac{\Lambda}{\Lambda^{\prime}}\right)
=\frac{2k^{\prime}\pi}{s},\quad k^{\prime}\in\mathbb{Z} \label{orthabapp2}%
\end{equation}
is the $\mathcal{A}_{0}$ term zero.

Now, for \eqref{liminfapp2} and its connection to the delta distribution, we
need to remember the Riemann-Lebesgue lemma
\begin{equation}
\lim_{u\rightarrow+\infty}\int_{\mathbb{R}}f(\omega)\cos\left(  \omega
u\right)  d\omega= 0
\end{equation}
and the representation of the Dirac delta
\begin{equation}
\lim_{u\rightarrow+\infty}\int_{\mathbb{R}}f(\omega)\frac{\sin\left(  \omega
u\right)  }{\omega}d\omega=\pi f(0)
\end{equation}
for an appropriate test function $f$. Then, \eqref{orthapp2} leads us to
\begin{equation}
\int_{0}^{+\infty}\!\!uW_{s,\sigma}W_{s,\sigma^{\prime}}du=\frac{1}{\sigma
}\delta(\sigma-\sigma^{\prime})+\frac{\mathcal{A}_{0}}{\sigma^{2}%
-\sigma^{\prime2}}%
\end{equation}

As we noticed before, the states that are orthogonal are characterized by
$\mathcal{A}_{0}=0$, which holds either for $s=0$ or under the validity of
\eqref{orthabapp2}. Otherwise, if $\mathcal{A}_{0}\neq0$, we notice that the
limit
\begin{equation}
\lim_{\sigma^{\prime}\rightarrow\sigma}\left(  \frac{\mathcal{A}_{0}}%
{\sigma^{2}-\sigma^{\prime2}}\right)  =\frac{s\tanh\left(  \frac{\pi s}%
{2}\right)  }{\pi\sigma^{2}}%
\end{equation}
is finite.

\end{document}